\documentclass[11pt]{article}
\usepackage{epsfig}
\newcommand{\la}[1]{\label{#1}}
\newcommand{\be}{\begin{equation}}
\newcommand{\ee}{\end{equation}}
\newcommand{\ba}{\begin{eqnarray}}
\newcommand{\ea}{\end{eqnarray}}
\newcommand{\bi}{\begin{itemize}}
\newcommand{\ei}{\end{itemize}}

\newcommand{\RR}{{\rm I\kern -.2em  R}}

\newcommand{\helio}{^6{\rm He}}
\newcommand{\neon}{^{18}{\rm Ne}}
\newcommand{\dis}{\nu_e\rightarrow\nu_e}
\newcommand{\disa}{\bar{\nu}_e\rightarrow\bar{\nu}_e}
\newcommand{\sigdma}{{\rm sign}(\Delta m^2_{23})}
\newcommand{\sigta}{{\rm sign}(\cos\theta_{23})}

\def\lsi{\raise0.3ex\hbox{$<$\kern-0.75em\raise-1.1ex\hbox{$\sim$}}}
\def\gsi{\raise0.3ex\hbox{$>$\kern-0.75em\raise-1.1ex\hbox{$\sim$}}}

\makeatletter \@addtoreset{equation}{section} \makeatother
\renewcommand{\theequation}{\arabic{section}.\arabic{equation}}
\makeatletter
\renewcommand\section{\@startsection {section}{1}{\z@}%
                                   {-5.5ex \@plus -1ex \@minus -.2ex}
                                   {2.3ex \@plus.2ex}%
                                   {\normalfont\large\bfseries}}
\renewcommand\subsection{\@startsection{subsection}{2}{\z@}%
                                     {-3.25ex\@plus -1ex \@minus -.2ex}%
                                     {1.5ex \@plus .2ex}%
                                     {\normalfont\normalsize\bfseries}}
\renewcommand\thesection {\@arabic\c@section}
\renewcommand\thesubsection   {\thesection.\@arabic\c@subsection}
\renewcommand{\@seccntformat}[1]{%
\csname the#1\endcsname.\hspace{1.0em}}
\makeatother

\begin{document}

\begin{titlepage}
\begin{flushright}
March 2005
\end{flushright}

IFIC/05-16      \\
FTUV-05-0303     \\

\begin{centering}

\vspace*{0.8cm}

\mbox{\Large\bf 
Optimal $\beta$-beam at the CERN-SPS
}

\vspace*{0.8cm}

J.~Burguet-Castell$^{\rm a,}$\footnote{jordi.burguet.castell@cern.ch},
D.~Casper$^{\rm b,}$\footnote{dcasper@uci.edu}, 
E.~Couce$^{\rm a,}$\footnote{ecouce@gmail.com}, 
J.J.~G\'omez-Cadenas$^{\rm a,}$\footnote{gomez@mail.cern.ch}, 
P.~Hern\'andez$^{\rm a,}$\footnote{pilar.hernandez@ific.uv.es}

\vspace*{0.8cm}

{\em $^{\rm a}$%
IFIC, 
Universidad de Val\`encia, 
E-46100 Burjassot, Spain\\}
\vspace{0.2cm}

{\em $^{\rm b}$%
Department of Physics and Astronomy, 
University of California, Irvine
CA 92697-4575, USA\\}

\vspace{1cm}

\mbox{\bf Abstract}

\end{centering}

A $\beta$-beam with maximum $\gamma=150$ (for $\helio$ ions) or 
$\gamma=250$ (for $\neon$) could be achieved at the CERN-SPS.
We study the sensitivity to $\theta_{13}$ and $\delta$ of such a beam 
as function of $\gamma$, optimizing with the baseline constrained
to CERN-Frejus (130~km), and also with simultaneous
variation of the baseline.  These results are compared to
 the {\it standard} scenario previously considered, with lower $\gamma=60/100$, and also 
with a higher $\gamma\sim 350$ option that requires a more powerful 
accelerator.  Although higher $\gamma$ is better, loss of sensitivity to $\theta_{13}$ and $\delta$ is most pronounced for $\gamma$ below 100.

\vfill

\end{titlepage}

\section{Introduction}

Results from atmospheric~\cite{atmos}, 
solar~\cite{solar}, reactor~\cite{reactor} and long-baseline~\cite{k2k} 
neutrino experiments in recent years can be economically accommodated
in the Standard Model~(SM) with neutrino masses and a three-neutrino mixing
matrix~\cite{MNS}. In this case, the lepton sector 
of the SM closely resembles that of the quarks and there are 
new physical parameters measurable at low energies: the three neutrino 
masses, $m_i$ ($i=1,2,3$), three mixing angles, $\theta_{ij}$, ($i\neq j = 1,2,3$),  and a CP-violating phase, $\delta$. In contrast 
with the quark sector, two additional phases could 
be present if neutrinos are Majorana. Of these new
parameters, present experiments have 
determined just two neutrino mass-square differences and two 
mixing angles: 
($|\Delta m^2_{23}| \simeq 2.2\times 10^{-3}~\hbox{eV}^2$, 
$\theta_{23}\simeq 45^\circ$) which mostly 
drive the atmospheric oscillation and 
($\Delta m^2_{12}\simeq 8\times 10^{-5}~\hbox{eV}^2$,
$\theta_{12}\simeq 32^\circ$) which mostly drive the solar one. 
The third angle, $\theta_{13}$,   as well as the 
CP-violating phases ($\delta$, and possible Majorana phases)
remain undetermined. Only an upper limit $\theta_{13} \leq 12^\circ$ is known.
Another essential piece of information needed to clarify the 
low-energy structure of the lepton flavor 
sector of the SM is the neutrino mass hierarchy and the absolute neutrino mass scale. The former is related to 
the sign of the largest mass-square difference ($\Delta m^2_{23}$), which 
determines if the spectrum is hierarchical 
(if the two most degenerate neutrinos are lighter than the third one) 
or degenerate (if they are heavier).

Measurement of some of these parameters may be possible in 
high-precision neutrino-oscillation experiments. A number
of experimental setups to significantly improve on present
sensitivity to $\theta_{13}$, $\delta$ and the sign of $\Delta m^2_{23}$ 
have been discussed in the literature:
neutrino factories (neutrino beams from boosted-muon 
decays)~\cite{geer,drgh,nufact}, 
superbeams (very intense conventional neutrino beams)~\cite{JHF,NUMI,splcern,superbeam}
improved reactor experiments~\cite{reactor_deg} 
and more recently $\beta$-beams (neutrinos 
from boosted-ion decays)~\cite{zucchelli,mauro}. These are 
quite different in terms of systematics but all 
face a fundamental problem which limits the 
reach of each individual experiment significantly, namely the correlations and degeneracies between 
parameters~\cite{golden}-\cite{silver};
$\theta_{13}$ and $\delta$  
must be measured simultaneously, and 
other oscillation parameters are not known with perfect precision.  

To resolve these degeneracies it is
important to measure as many independent channels as possible and to 
exploit the energy and/or baseline dependence of the 
oscillation signals and matter effects in neutrino propagation. In many 
cases, the best way to do this is by combining different experiments; 
indeed the synergies between some combinations of the setups mentioned above  
have been shown to be considerable.

The neutrino factory provides ultimate sensitivity to leptonic CP violation,
and thus represents the last step on a long-term road map to reveal 
the lepton-flavor sector of the SM. 
Recently it was shown that a $\beta$-beam
running at a higher $\gamma$ than previously considered (and longer baselines),
in combination with a massive water detector, 
can reach sensitivity to leptonic CP-violation 
and $\sigdma$ that competes with a neutrino 
factory's.  The optimal setup among those considered in~\cite{bbeam} was a
$\beta$-beam with $\gamma=350/580$ for $\helio$ and $\neon$ isotopes
respectively and a baseline $L\simeq 730$~km. If constructed at CERN, this beam would require 
a refurbished SPS or an acceleration scheme utilizing the LHC - implying substantial R\&D effort in either case.

This paper considers instead the possibility of using the existing CERN-SPS
up to its maximum power, allowing a beam with $\gamma=150\,(250)$ for $\helio\,(\neon)$ ions (some preliminary results of this study were presented in \cite{now04}). The design of this $\beta$-beam is essentially as described in~\cite{bbcern}.

The advantages of increasing the $\gamma$ factor discussed in~\cite{bbeam} 
also apply in this case.
The oscillation signals grow at least linearly with the $\gamma$ factor, 
therefore the highest $\gamma$ possible is preferred in principle, 
if the baseline is adjusted appropriately. Furthermore when the 
energy is
well above the Fermi momentum of the target nuclei, energy dependence of the oscillation signals
is very effective in resolving parameter degeneracies.  In practice there are
two caveats to this rule.  First, water Cherenkov detectors are best suited   
for quasi-elastic (QE) reactions, where the neutrino 
energy can be kinematically reconstructed.  Therefore sensitivity improves with $\gamma$ 
only until the inelastic cross section begins to dominate; we will show that this
occurs for $\gamma \geq 400$. 
The second concern is 
background, since NC single-pion production can mimic the
the appearance signal; it is demonstrated in~\cite{bbeam} and confirmed here that this
background is manageable, even for $\gamma > 100$.

More concretely the purpose of this paper is two-fold. First,
optimization of a CERN-SPS $\beta$-beam by answering the following questions:
\begin{itemize}
\item Assuming an underground laboratory at Frejus with a
 megaton water Cherenkov detector, what is the optimal $\gamma$ with the existing 
CERN-SPS?
\item For the maximum $\gamma$ achievable with the CERN-SPS, what is the optimal $\beta$-beam baseline?
\item Is there any physics advantage to varying the $\gamma$ ratio for $\helio$ and
$\neon$, i.e. a ratio different from $\gamma_{\neon}/\gamma_{\helio} = 1.67$ (which allows both beams to circulate simultaneously)?~\cite{matsmo}
\end{itemize}
Second, comparing the performance of the following set-ups:
\begin{itemize}
\item Setup I: $L=130$~km (CERN-Frejus) at the optimal $\gamma$ accessible to 
the CERN-SPS.
\item Setup II: $\gamma =150$ at the optimal baseline.
\item Setup III: $\gamma=350$ at
$L=730$~km, which is a symmetric version of the configuration considered in~\cite{bbeam}. To accelerate the ions would require either
a refurbished SPS (with superconducting magnets) or a more powerful
accelerator, such as the Tevatron or LHC. 
\end{itemize}

In all cases an intensity of $2.9\times 10^{18}~\helio$ and 
$1.1\times 10^{18}~\neon$ decays per year~\cite{bbcern} and an
integrated luminosity corresponding to 10~years are assumed. Although these luminosities have been estimated for simultaneous ion circulation (fixing the ratio of $\gamma$'s to $1.67$) reference~\cite{matspc} argues they are achievable even if the ions circulate separately at the same $\gamma$, by injecting more bunches.  While these intensities are realistic for the CERN-SPS, the same has not been demonstrated for other accelerators like the Tevatron or LHC.  The far detector is a Super-Kamiokande-like
water Cherenkov design, with fiducial mass 440~kton.

The paper is organized as follows. Section~2 shows
expected fluxes and event rates  for the maximum $\gamma$ achievable at the CERN-SPS. Section~3 describes the performance of a large water Cherenkov detector for the appearance and disappearance signals and estimates the atmospheric background, an important constraint in design of the bunch length. Section~4 deals with optimizations needed to define setups I and II and Section~5
compares the physics reach of the three emergent reference setups. Section~6 discusses our outlook and conclusions.

\section{Neutrino fluxes and rates}

Figure~\ref{fig:fluxes} shows the fluxes for the maximum acceleration of the ions at the CERN-SPS: $\gamma=150$ for $\helio$ and $\gamma=250$ for $\neon$ at $L=300$~km. Table~1
shows the rate of charged-current interactions expected per kiloton in one year.

\begin{figure}[t]
\begin{center}

\epsfig{file=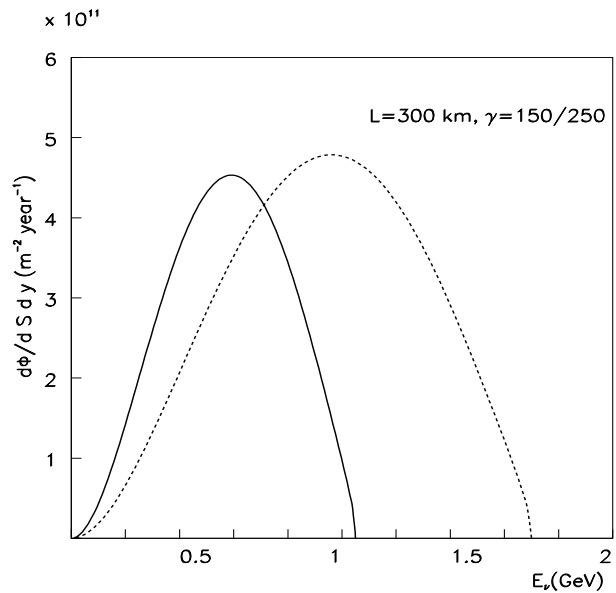,width=9cm,height=8cm} 
\caption[a]{\it $\bar{\nu}_e$ (solid) and $\nu_e$ (dashed) fluxes as a function of the neutrino energy at $L=300$~km for the maximum acceleration of the $\helio$ ($\gamma=150$) and $\neon$ ($\gamma=250$) at the CERN-SPS.}
\la{fig:fluxes}

\end{center}
\end{figure}

\begin{table}
\begin{center}
\begin{tabular}{|c|c|c|c|c|}
\hline
$\gamma$ & $L(km)$ & ${\bar\nu}_e$ CC & $\nu_e$ CC & $\langle E_\nu \rangle (GeV)$\\
\hline
\hline
150/250 & 300 & 22.8 & 115.6 & 0.58/0.94 \\
\hline
\hline
\end{tabular}
\label{table:CCevents}
\caption{\it Number of charged-current events per \hbox{kton-year}, in the absence of oscillation,
for the maximum acceleration of $\helio$ and $\neon$ at the CERN-SPS.  The average neutrino energy is also shown.}
\end{center}
\end{table}

\section{Measurements at a $\beta$-beam}

The parameters $\theta_{13}$ and $\delta$ are best studied by probing the appearance channels for neutrino oscillation in the atmospheric energy range: golden ($\nu_\mu \leftrightarrow \nu_e$)~\cite{drgh,golden} and silver 
($\nu_\tau \leftrightarrow \nu_e$)~\cite{silver} channels have been identified. In the setups considered here, neutrino energies are below $\tau$ threshold, therefore only the golden channel is available. 

The disappearance transition $\nu_e \rightarrow \nu_e$ can also be measured. This is an important complement to the golden channel measurement, because the intrinsic degeneracy~\cite{burguet1} 
in the golden measurement can be resolved: the disappearance measurement depends on $\theta_{13}$, but not on 
$\delta$.  The synergy between the appearance and disappearance channels for a
$\beta$-beam is thus analogous to that between superbeam and reactor experiments~\cite{reactor_deg}.


\subsection{Detection of the appearance signal}
\label{detectors}

The signal for the golden transition is a charged-current event (CC) with a muon in the final state. 
Reference~\cite{bbeam} studied the performance of a 440~kton fiducial water Cherenkov detector similar to Hyper-Kamiokande or the proposed by the UNO experiment~\cite{uno}.  This analysis can be extended to
different $\gamma$'s, using the same neutrino
physics generator, detector simulation and reconstruction
algorithms as described in~\cite{bbeam}, with realistic e-$\mu$ separation by pattern recognition, and the requirement of a
delayed coincidence from muon decay.

\begin{figure}[t]
\epsfig{file=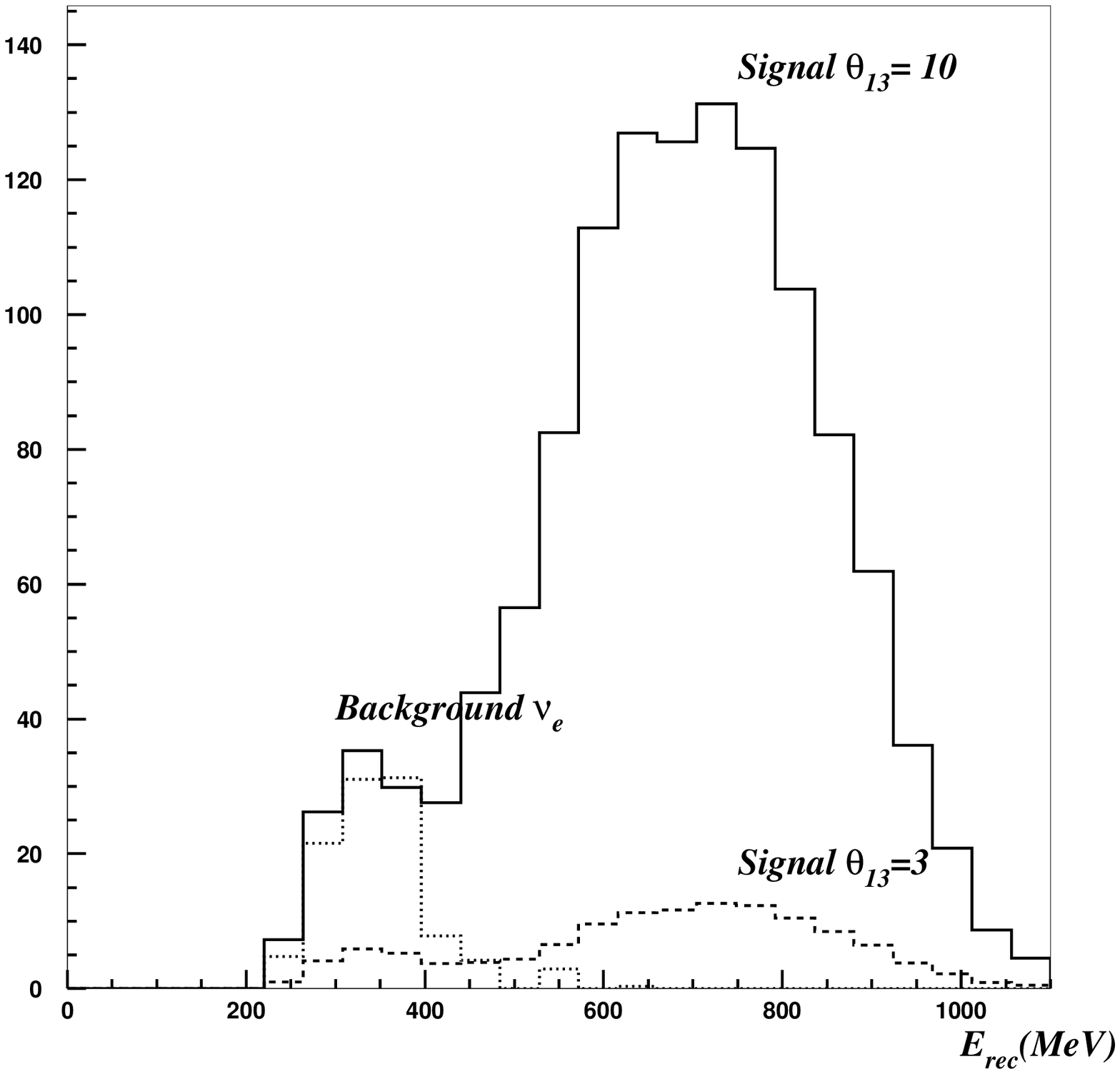,width=7.5cm,height=7.5cm} 
\epsfig{file=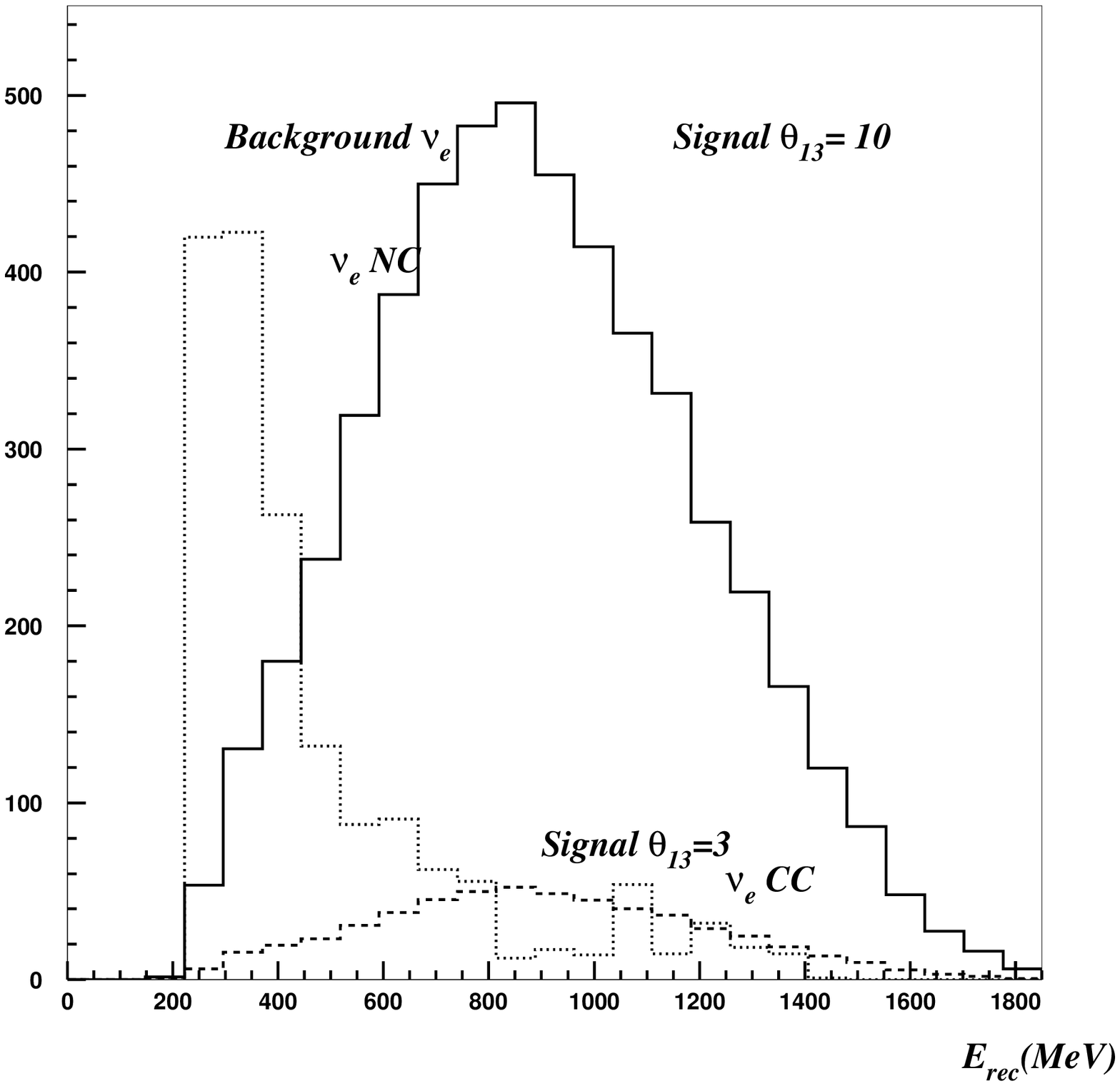,width=7.5cm,height=7.5cm} 

\caption{\it Reconstructed energy for signal with $\theta_{13}=8^\circ$ (solid) and $\theta_{13}=3^\circ$ (dashed)) and background (dotted) at the maximum 
acceleration of $\helio$ (left) and $\neon$ (right) ions at the CERN-SPS. The absolute normalization corresponds to one year.}
\la{fig:erec}

\end{figure}
Figure~\ref{fig:erec} shows the reconstructed energy spectra of
signal and background at maximum CERN-SPS $\gamma$, for two different values of $\theta_{13}$.
Backgrounds are smaller for $^{6}$He than $^{18}$Ne, and both neutrino and anti-neutrino
backgrounds tend to cluster at low energies. Most of the background reconstructs below
$500$~MeV. 

\begin{figure}[t]
\epsfig{file=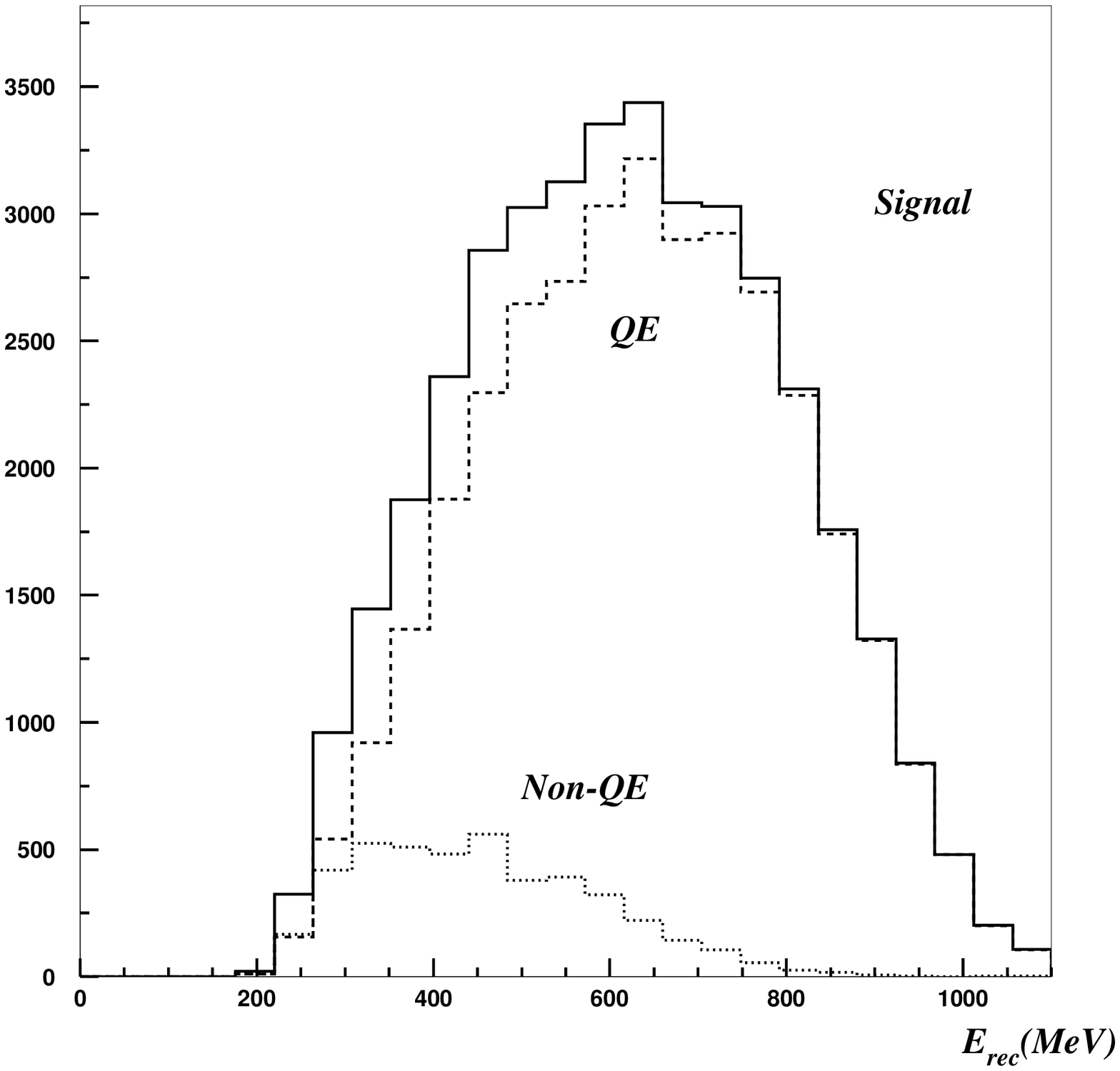,width=7.5cm,height=7.5cm} 
\epsfig{file=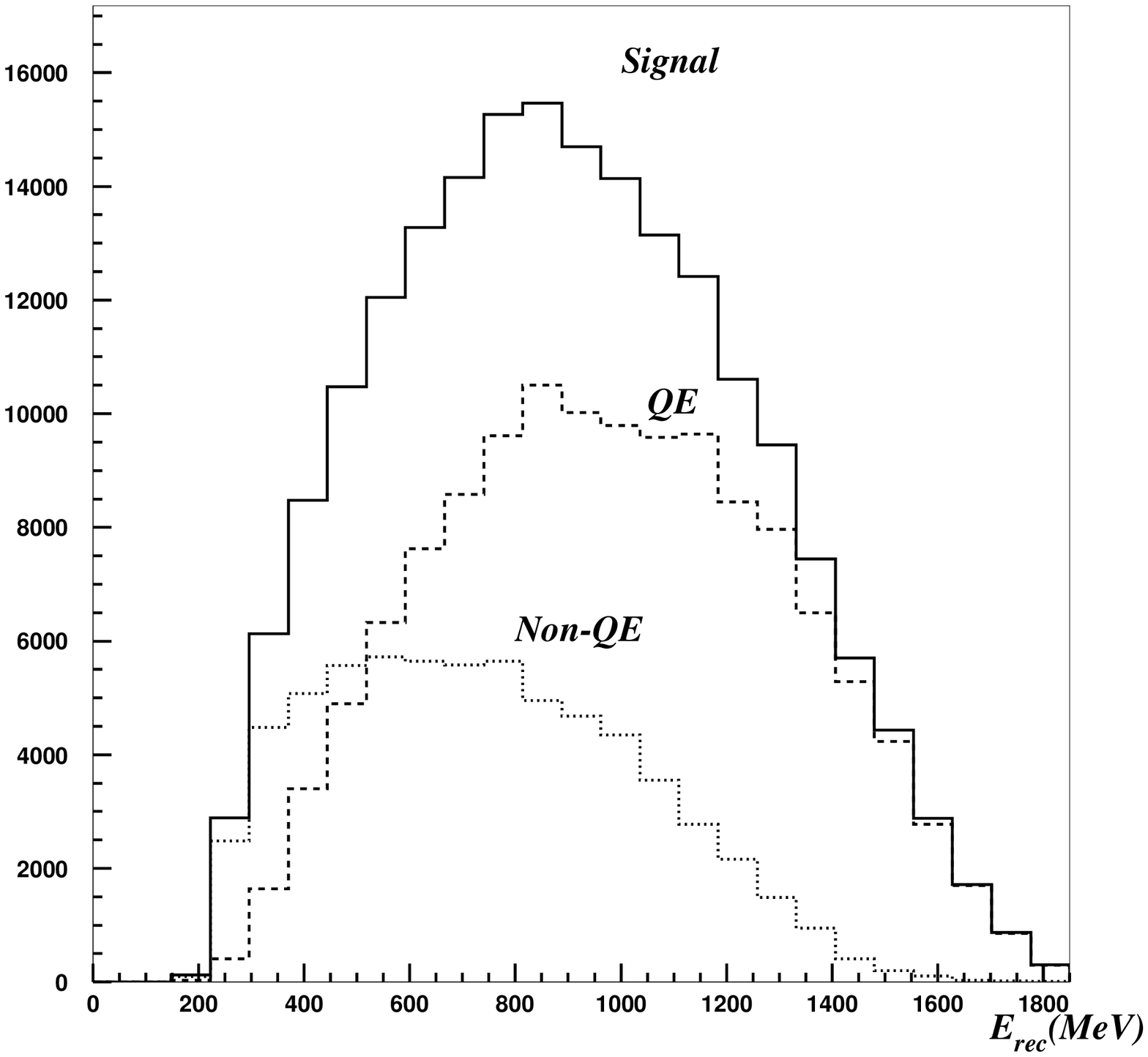,width=7.5cm,height=7.5cm} 

\caption{\it Quasi-elastic and non-quasielastic components in the $\mu$ appearance signal for unit oscillation probability (the absolute normalization is arbitrary) at maximum CERN-SPS acceleration of $\helio$ (left) and $\neon$ (right).}
\la{fig:qevsnqe}
\end{figure}

The neutrino energy resolution depends strongly on the proportion of quasi-elastic (QE) 
and non quasi-elastic (non-QE) interactions in the signal.  Neutrino energy is reconstructed
assuming two-body, quasi-elastic kinematics, so contamination from non-QE events introduces a bias 
between the true and reconstructed energies.
Figure~\ref{fig:qevsnqe} shows the fraction of QE and non-QE events passing the
selection criteria.
As expected the non-QE contamination is smaller for anti-neutrinos since 
the average beam energy is also smaller for the chosen $\gamma$'s. 

To properly include both detector resolution and non-QE contamination effects, a matrix describing the migration between true and reconstructed neutrino energies is constructed. Migration matrices are also computed for the backgrounds. Given the irreducible Fermi motion and muon threshold, the first energy bin extends from 0-500~MeV and bins of 250~MeV 
width are used above it. For the high-$\gamma$ Setup III, the first bin is discarded.

Table~\ref{tab:effbgapp} shows these migration matrices for $\gamma=120$ and 150 for $\helio$ and $\neon$.
\footnote{Matrices with appropriate binning for other choices of $\gamma$ can be obtained from the authors on request.}. 
Three bins: 0-500, 500-750 and 750+~MeV are used.  The efficiencies are quite high 
($\sim 30-50\%$) even when the background fraction is held below $10^{-3}$.

\begin{table}
\begin{center}
\(
\begin{array}{|c|c|c|c|}
\hline
{\rm Ion} & \gamma & \epsilon^{app}_{ij} & b^{app}_{ij} \\
\hline
\hline
\helio & 120 & \begin{footnotesize}\left(\begin{array}{lll}
0.65&0.18&0.071\\
0.03&0.54&0.33\\
0.&0.016&0.34\\
\end{array}\right) \end{footnotesize} &  \begin{footnotesize}\left(\begin{array}{lll}
0.21\times 10^{-3}&0.30\times 10^{-2}&0.25\times 10^{-2}\\
0.&0.68\times 10^{-4}&0.20\times 10^{-3}\\
0.&0.&0.\\
\end{array}\right) \end{footnotesize} \\
\hline
\neon & 120 &  
\begin{footnotesize}
\left(\begin{array}{lll}
0.47&0.18&0.11\\
0.050&0.34&0.23\\
0.77\times10^{-3}&0.30\times10^{-1}&0.14\\
\end{array}\right) \end{footnotesize} &  \begin{footnotesize}\left(\begin{array}{lll}
0.73\times 10^{-3}&0.20\times 10^{-2}&0.30\times 10^{-2}\\
0.12\times 10^{-3}&0.55\times 10^{-3}&0.11\times 10^{-2}\\
0.& 0.39\times 10^{-4}& 0.74\times 10^{-3}\\
\end{array}\right) \end{footnotesize} \\
\hline
\helio & 150 & \begin{footnotesize}\left(\begin{array}{lll}
0.66&0.15&0.056\\
0.034&0.56&0.20\\
0.&0.029&0.44 \\
\end{array}\right) \end{footnotesize} &  \begin{footnotesize}\left(\begin{array}{lll}
0.22\times 10^{-3}&0.31\times 10^{-2}&0.24\times 10^{-2}\\
0.&0.80\times 10^{-4}&0.12\times 10^{-3}\\
0.& 0.&0.\\
\end{array}\right) \end{footnotesize} \\
\hline
\neon & 150 &  
\begin{footnotesize}\left(\begin{array}{lll}
0.47&0.16&0.082\\
0.054&0.34&0.16\\
0.84\times 10^{-3}&0.04&0.23\\
\end{array}\right) \end{footnotesize} &  \begin{footnotesize}\left(\begin{array}{lll}
0.78\times 10^{-3}&0.22\times 10^{-2}&0.35\times 10^{-2}\\
0.12\times 10^{-3}&0.66\times 10^{-3}&0.64\times 10^{-3}\\
0.&0.47\times 10^{-4}&0.80\times 10^{-3}\\
\end{array}\right) \end{footnotesize} \\
\hline
\hline
\end{array}\)
\caption{\it Migration matrices for appearance signal ($\epsilon_{ij}^{app}$) and backgrounds ($b_{ij}^{app}$) at different values of $\gamma$.  Each row and column of the matrices corresponds to a neutrino energy bin, as described in the text.}
\label{tab:effbgapp}
\end{center}
\end{table}


%
%






\subsection{Detection of the disappearance signal}

For $\dis$($\disa$) transitions, the signal is a CC interaction
with an electron(positron) in the final state. In~\cite{bbeam} this channel was included with a conservatively estimated 50\% flat efficiency and negligible background. 
Since the energy resolution is also strongly affected by the non-QE contamination for this sample,
this analysis is now refined to include the effect of migrations.  While the background level for this large signal can be safely neglected in comparison to other systematic errors to be discussed later, a matrix of efficiencies should be used to account for the signal migrations. Table~\ref{tab:effdis} shows these matrices for $\helio$ and $\neon$ at various $\gamma$'s. Efficiencies are quite high, especially at lower energies where they reach 80-90\%. 

\begin{table}
\begin{center}
\(
\begin{array}{|c|c|c|}
\hline
Ion & \gamma & \epsilon^{dis}_{ij} \\
\hline
\hline
\helio & 120 & \begin{footnotesize}\left(\begin{array}{lll}
0.89&0.25&0.10\\
0.04&0.62&0.40\\
0.&0.023&0.38\\
\end{array}\right) \end{footnotesize} \\
\hline
\neon & 120 &  
\begin{footnotesize}\left(\begin{array}{lll}
0.83&0.35&0.21\\
0.073&0.46&0.36\\
0.15\times 10^{-2}&0.43\times 10^{-1}&0.22\\
\end{array}\right) \end{footnotesize} \\
\hline
\helio & 150 & \begin{footnotesize}\left(\begin{array}{lll}
0.89&0.21&0.086\\
0.045&0.63&0.25\\
0.&0.041&0.52\\
\end{array}\right) 
\end{footnotesize} \\
\hline
\neon & 150 &  
\begin{footnotesize}\left(\begin{array}{lll}
0.83&0.33&0.16\\
0.078&0.47&0.27\\
0.19\times 10^{-2}&0.059&0.33\\
\end{array}\right) 
\end{footnotesize} \\
\hline
\hline
\end{array}
\)
\caption{\it Fractional migration matrices ($\epsilon_{ij}^{dis}$) of the CC $\nu_e$ disappearance signal for different values of $\gamma$.}
\label{tab:effdis}
\end{center}
\end{table}





\subsection{Atmospheric background}

\begin{figure}[bh]
\begin{center}
\epsfig{file=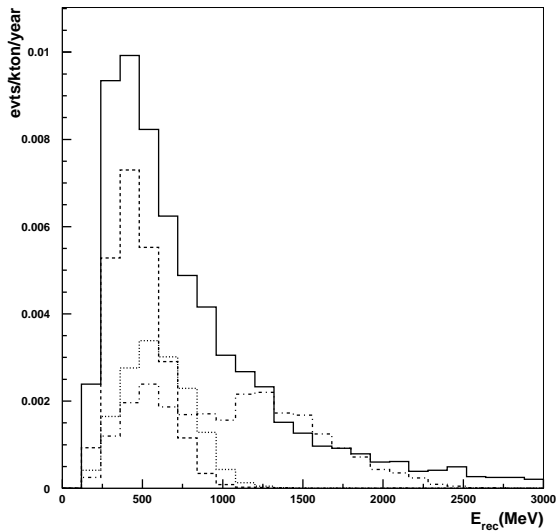,width=8cm}
\end{center}
\caption[]{\it Solid line: energy spectrum 
of atmospheric $\nu_{\mu}+\bar{\nu}_{\mu}$
background per kiloton-year, scaled down by a factor 1/500. 
Dashed, dotted
and dash-dotted lines: energy spectrum of signal events per kiloton-year for $\gamma=120, 150$ and $350$ assuming $\theta_{13} = 1^\circ$. }
\label{fig:atm1}
\end{figure}

An important background for any accelerator-based experiment to
control arises from atmospheric neutrinos. A detector like Super--Kamiokande
will expect approximately 120 $\nu_{\mu}+\bar{\nu}_{\mu}$ interactions per kiloton-year
(including the disappearance of $\nu_{\mu}$ into $\nu_{\tau}$). 
Of these, 32 atmospheric $\nu_{\mu}+\bar{\nu}_{\mu}$ per kiloton-year 
pass all the selection cuts (one non-showering ring, accompanied by a delayed coincidence from
muon decay). The reconstructed spectrum of those events {\it scaled by a factor 1/500}
is shown in Figure~\ref{fig:atm1} (solid line) alongside the signal for the three example
setups to be considered later, namely,  
$\gamma=120$ ($L=130$ km, dashed),  $\gamma=150$ ($L=300$ km, dotted)
and $\gamma=350$  $L=730$ km, dashed-dotted), assuming $\theta_{13}= 1^\circ$.

There are two additional handles to further reduce the atmospheric background. First, at a given $\gamma$,
we know the end-point of the signal spectrum, and there is no efficiency penalty for excluding events above the maximum beam energy. 
This cut obviously works best for lower-$\gamma$ scenarios.  Table~\ref{tab:cosmic} shows the effect of the 
end-point cut for different $\gamma$'s. For higher $\gamma$, it is also helpful to set a {\it lower} energy cut.  Requiring $E\geq 500$~MeV, for instance, is free for the highest $\gamma=350$ option, since this bin is not considered in the analysis anyway. 
\begin{table}
\begin{center}
\begin{tabular}{|c|c|c|c|c|}
\hline
$\gamma$ & Selection & $E_{max}$ cut & $E_{min}$ cut & $\cos\theta_l$ cut\\
\hline
\hline
120 & 32 & 19 & 19 & 15\\
\hline
150 & 32 & 24 & 24 & 15 \\
\hline
350 & 32 & 30 & 19 & 5 \\
\hline
\hline
\end{tabular}
\caption{\it Surviving atmospheric $\nu_{\mu}$ background per kton-year after
cuts: on the high-energy end-point of the 
$\beta$-beam neutrino spectrum ($E_{max}$), the low-energy tail ($E_{min}$) for setup III, and the lepton scattering angle ($\cos\theta_l$), as described in the text.}
\label{tab:cosmic}
\end{center}
\end{table}

\begin{figure}[bh]
\begin{center}
\epsfig{file=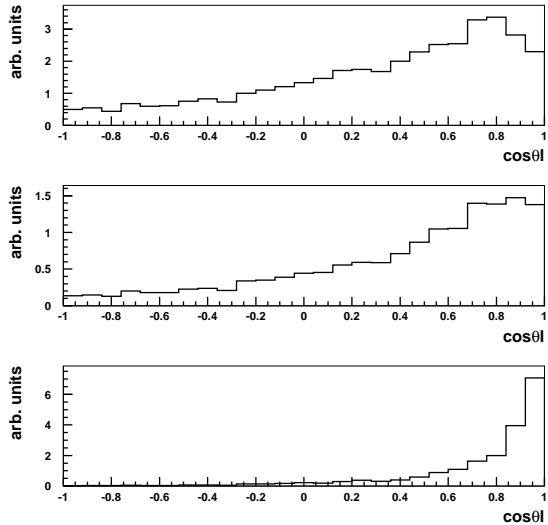,width=8cm}
\end{center}
\caption[]{\it Cosine of the reconstructed neutrino-lepton scattering angle for three setups: $\gamma=120$ (top),
$\gamma=150$ (middle) and $\gamma=350$ (bottom).} 
\label{fig:atm2}
\end{figure}

Second, a directional cut is also possible, since the beam arrives from a specific, known direction but the atmospheric background is roughly isotropic.  While the neutrino direction cannot be measured directly, it is increasingly correlated with the observable lepton direction at high energies.  Figure~\ref{fig:atm2} illustrates this correlation for the three reference set-ups.  Thus, a directional cut is more effective as $\gamma$ increases, but is never perfectly efficient.  To compare the 
power of this cut for the different setups, we define it to achieve a 90\% efficiency in all cases: $\cos \theta_{l} > 0.45$ for $\gamma=350$, $\cos \theta_{l} > -0.3$ for $\gamma=150$ and $\cos \theta_{l} > -0.5$ for $\gamma=120$.  The remaining atmospheric background for each setup is summarized in Table~\ref{tab:cosmic}.  Thanks to the directional cut, background rejection for the highest $\gamma$ is a factor three better than the alternative scenarios. 

Even with energy and directional cuts, 5 to 15 atmospheric $\nu_{\mu}$ background events per kiloton-year remain, compared
to the expected intrinsic beam-induced detector background (mostly due to NC single-pion production) of ${\cal O}(10^{-2})$ events. 
To reduce atmospheric contamination to a negligible level (say ten times below the intrinsic background) would require a rejection factor ${\cal O}(10^4)$ , although since the atmospheric background can be well measured a rejection factor 5-10 times less stringent is probably tolerable.

This rejection factor can be achieved by timing of the parent ion bunches. It is estimated~\cite{mauro} that 
a rejection factor of $2 \times 10^4$ is feasible with bunches 10~ns in length.  Based on the present results, a less demanding scheme for the number of bunches and bunch length could be workable.

\subsection{Systematic errors}

Although a detailed analysis of all possible systematic errors is beyond the scope 
of this paper, we have included the two that will likely dominate.
First, the uncertainty in the fiducial mass of the near and far detectors, which we estimate as a $\pm 5\%$ effect on the expected far-detector rate.  Second, the uncertainty on the ratio of anti-neutrino/neutrino cross sections, which we assume a near detector can measure with an accuracy of $\pm 1\%$\footnote{The calculable neutrino and anti-neutrino energy spectra of the $\beta$-beam will facilitate cross-section measurements, compared to a traditional neutrino beam.}.

To include these errors, two new parameters are added to the fits: $A$, the global normalization,
and $x$, the relative normalization of anti-neutrino to neutrino rates.  More precisely, if $n^{i,\pm}_{\mu,e}$ 
is the number of {\it measured} muon and electron events in the energy bin $i$
for the anti-neutrino ($+$) or neutrino ($-$) beam, and $N^{i,\pm}_{\mu,e}(\theta_{13},\delta)$ is the 
expected number for some values of the unknown parameters $(\theta_{13},\delta)$, then we minimize the following $\chi^2$ function:
\ba
\chi^2(\theta_{13},\delta,A,x)= 2 \sum_{i,f=e,\mu} \left\{A x N^{i,+}_f - n^{i,+}_f + n^{i,+}_f \log\left({n^{i,+}_f \over A x N^{i,+}_f}\right) \right. \nonumber \\
+ \left. A N^{i,-}_f - n^{i,-}_f + n^{i,-}_f \log\left({n^{i,-}_f \over A N^{i,-}_f}\right) \right\} + {(A-1)^2 \over \sigma_A^2} +{(x-1)^2 \over \sigma_x^2}.
\ea
where $\sigma_A=0.05$ and $\sigma_x=0.01$. The minimization in the parameters $A$ and $x$ for fixed values of $\theta_{13}$ and $\delta$ can be done analytically to leading order in the deviations 
$A-1$ and  $x-1$, that is solving the linearized system:
\ba
{\partial \chi^2 \over \partial A} =0 \;\;\;\;{\partial \chi^2 \over \partial x} =0
\ea

In what follows, sensitivity to the parameters $(\theta_{13}, \delta)$ will be quantified 
using 99\% confidence regions for two 
degrees of freedom; that is, the curves satisfying:
\ba
\chi^2(\theta_{13},\delta,A_{min},x_{min}) = 9.21.
\ea

\section{Optimization of the CERN-SPS $\beta$-beam }

The following sensitivity plots are used to optimize the physics performance of different $\beta$-beams:

\begin{itemize}
\item Sensitivity to CP violation: region on the plane $(\theta_{13},\delta)$ where the phase $\delta$ can be distinguished from both $\delta=0^\circ$ and $\delta=180^\circ$ for {\it any}
best fit value of $\theta_{13}$, at 99\% confidence level or better.
\item Sensitivity to $\theta_{13}$: region on the plane $(\theta_{13},\delta)$ where the angle $\theta_{13}$ can be distinguished from $\theta_{13}=0$ for any best fit value of $\delta$, at 99\% confidence level or better.
\end{itemize}

Unless otherwise specified, the following solar- and atmospheric-neutrino oscillation parameters are assumed:
\ba
\Delta m^2_{12}= 8.2\times 10^{-5} eV^2 \;\;\;\; \theta_{12}=32^\circ\;\;\;\;
\Delta m^2_{23}= 2.2\times 10^{-3} eV^2 \;\;\;\; \theta_{23}=45^\circ
\ea

\subsection{Optimal $\gamma$ for the CERN-Frejus baseline}

One frequently considered {\it standard} setup adopts the CERN--Frejus baseline $L=130$~km and $\gamma = 60/100$ for
$\helio$/$\neon$~\cite{mauro,blm}. This setup appears to be far from optimal even if the baseline is kept fixed.  As noted in~\cite{bbeam}, a higher-$\gamma$ beam increases the event rate and allows the energy dependence of the signal to be analyzed.  Taking 
the identical $\gamma$ for $\helio$ and $\neon$, Figure~\ref{fig:scan_l130} shows the $\gamma$-dependence of the 99\%~CL $\delta$ and $\theta_{13}$ sensitivity, as defined above.  The stars indicate the values of the previously considered setup in~\cite{mauro,blm}, corresponding to $\gamma=60/100$.  Clearly the CP-violation sensitivity is significantly better for larger $\gamma$. For
$\gamma \geq 100$ the sensitivity to CP violation and $\theta_{13}$ changes rather slowly.  This is not surprising, since increasing $\gamma$ at fixed baseline does not reduce the flux significantly at low energies (see Figure~\ref{fig:scaling}), just as for a Neutrino Factory.  In the absence of backgrounds, there is no penalty associated with higher $\gamma$,  although in practice, the non-negligible backgrounds result in a small decrease in $\theta_{13}$ sensitivity at higher $\gamma$, for some values
of $\delta$.

\begin{figure}[t]
\begin{center}

\epsfig{file=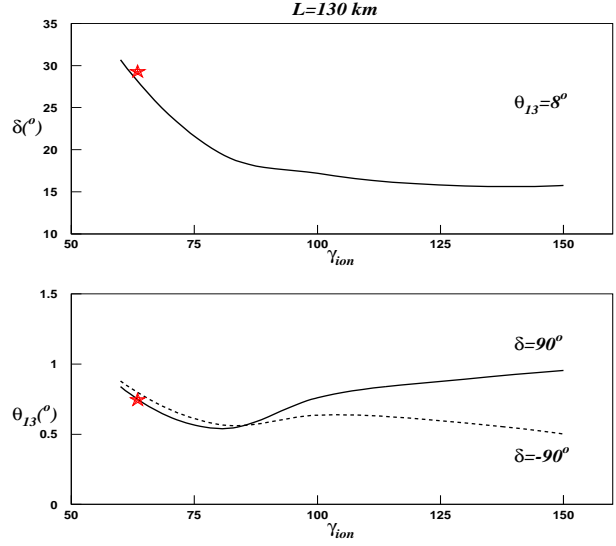,width=9cm,height=8cm} 

\caption{\it $\gamma$-dependence of 99\% confidence level $\delta$ sensitivity at $\theta_{13} = 8^\circ$ (top) and $\theta_{13}$ sensitivity (bottom) for $\delta = +90^\circ$ (solid) and $\delta = -90^\circ$ (dashed), assuming $L = 130$~km and $\gamma_{\helio} = \gamma_{\neon}$.  The stars indicate the values for the $\gamma = 60/100$ option in~\cite{mauro,blm}.}
\la{fig:scan_l130}

\end{center}
\end{figure}

\begin{figure}[t]
\begin{center}

\epsfig{file=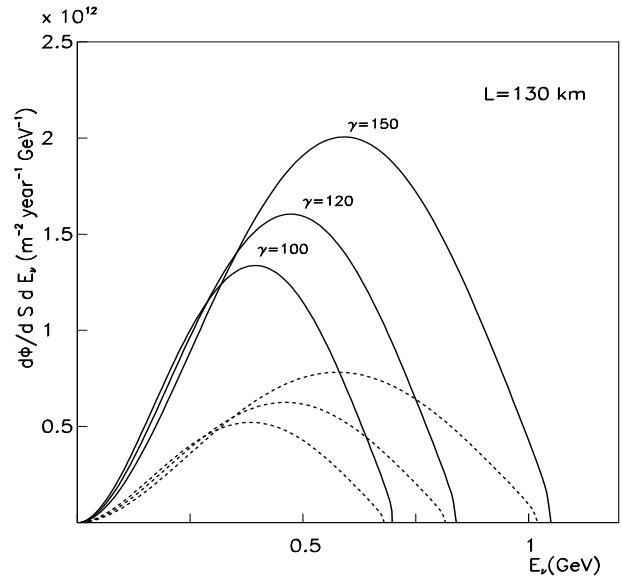,width=9cm,height=8cm} 

\caption{\it Energy spectra of $\nu_e$ (dashed) and $\bar{\nu}_e$(solid) at $L = 130$~km for $\gamma=100,120,150$.}
\la{fig:scaling}

\end{center}
\end{figure}

Although there is no unique optimal $\gamma$ within the wide range $\gamma=100-150$ when the baseline is fixed to $L=130$~km, consider for illustration an intermediate $\gamma=120$ to define Setup I; a different choice of $\gamma > 100$ will not make a significant difference.

There appears to be no advantage to the asymmetric choice 
$\gamma_{\neon}/\gamma_{\helio} =1.67$. The asymmetric option is always 
comparable in sensitivity to a symmetric one with the smaller $\gamma$ of the two, 
so a symmetric $\gamma$ configuration is adopted for setup I.

\subsection{Optimal $L$ for maximum ion acceleration $\gamma=150$}

As argued in~\cite{bbeam}, physics performance should 
improve with increasing $\gamma$, if the baseline is correspondingly scaled
to remain close to the atmospheric oscillation maximum, due to the (at least) linear increase in rate with $\gamma$.
This growth in sensitivity eventually saturates for a water detector, which becomes inefficient in reconstructing neutrino
energies in the inelastic regime.  Figure~\ref{fig:sat}, where the number of CC appearance candidates selected (for unit oscillation probability) is plotted as a function of $\gamma$ (for $\gamma/L$ fixed), confirms this expectation.  Saturation occurs for 
$\gamma \simeq 400$, above the maximum acceleration possible at the CERN-SPS, since the flux is still large
in the quasi-elastic region (see Figure~\ref{fig:scaling}). 

\begin{figure}[t]
\begin{center}

\epsfig{file=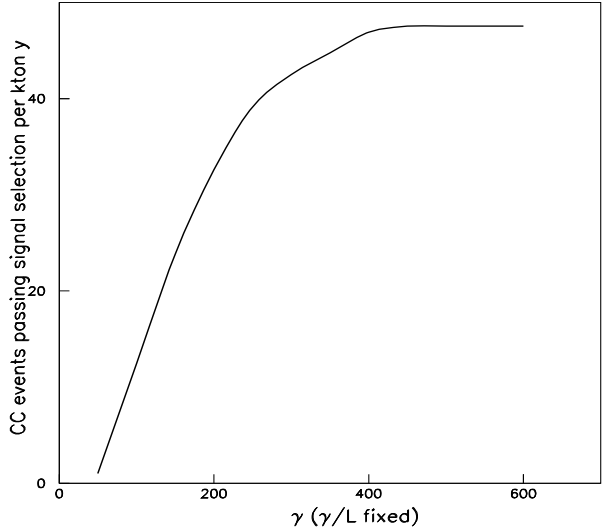,width=9cm,height=8cm} 

\caption{\it Number of CC appearance candidates (from $\neon$) for unit oscillation probability, as a function of $\gamma$, holding $\gamma/L$ fixed.} 
\la{fig:sat}

\end{center}
\end{figure}

Fixing $\gamma$ to the CERN-SPS we next study  the optimal baseline
and how the symmetric $\gamma$ setup compares with the asymmetric one. 

Figure~\ref{fig:versusL} shows the $|\delta|$ and $\theta_{13}$ sensitivities as a function of 
the baseline for $\gamma=150/150$ and the asymmetric case $\gamma=150/250$. 
The best CP sensitivity is achieved around $L\simeq 300(350)$~km for symmetric(asymmetric) beams \footnote{
The optimal baseline will obviously shift if $\Delta m^2_{23}$ is varied from the present best fit value: $\pm 50$ km for a change of one $\sigma$.}. The baseline dependence of $\theta_{13}$ sensitivity leads to similar conclusions, although the importance of choosing the optimum baseline is more pronounced.  A significant loss of $\theta_{13}$ sensitivity results if the baseline is too short, 
as in Setup I. 

\begin{figure}[t]
\begin{center}

\epsfig{file=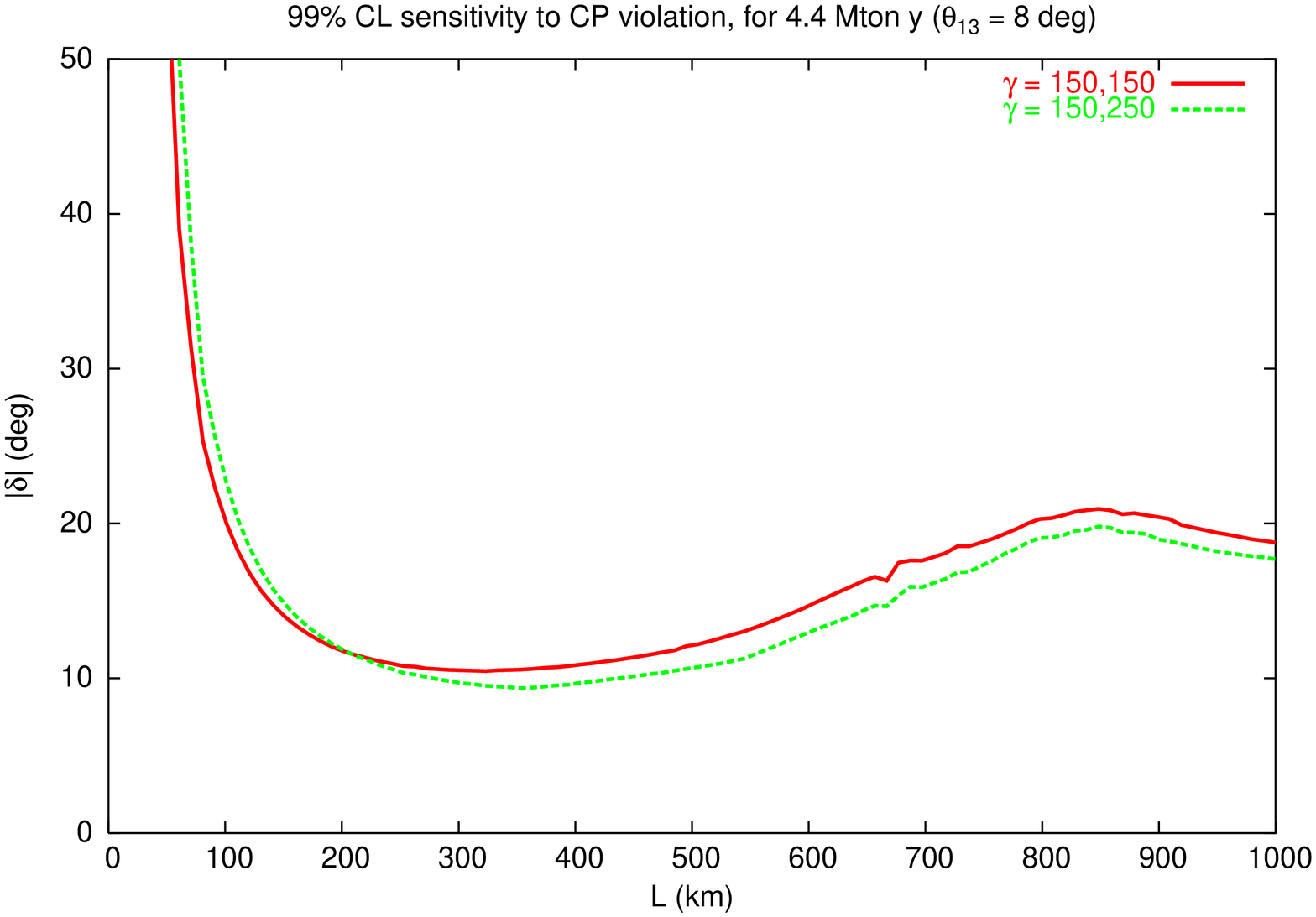,width=7cm,height=7cm,angle=-90} 
\epsfig{file=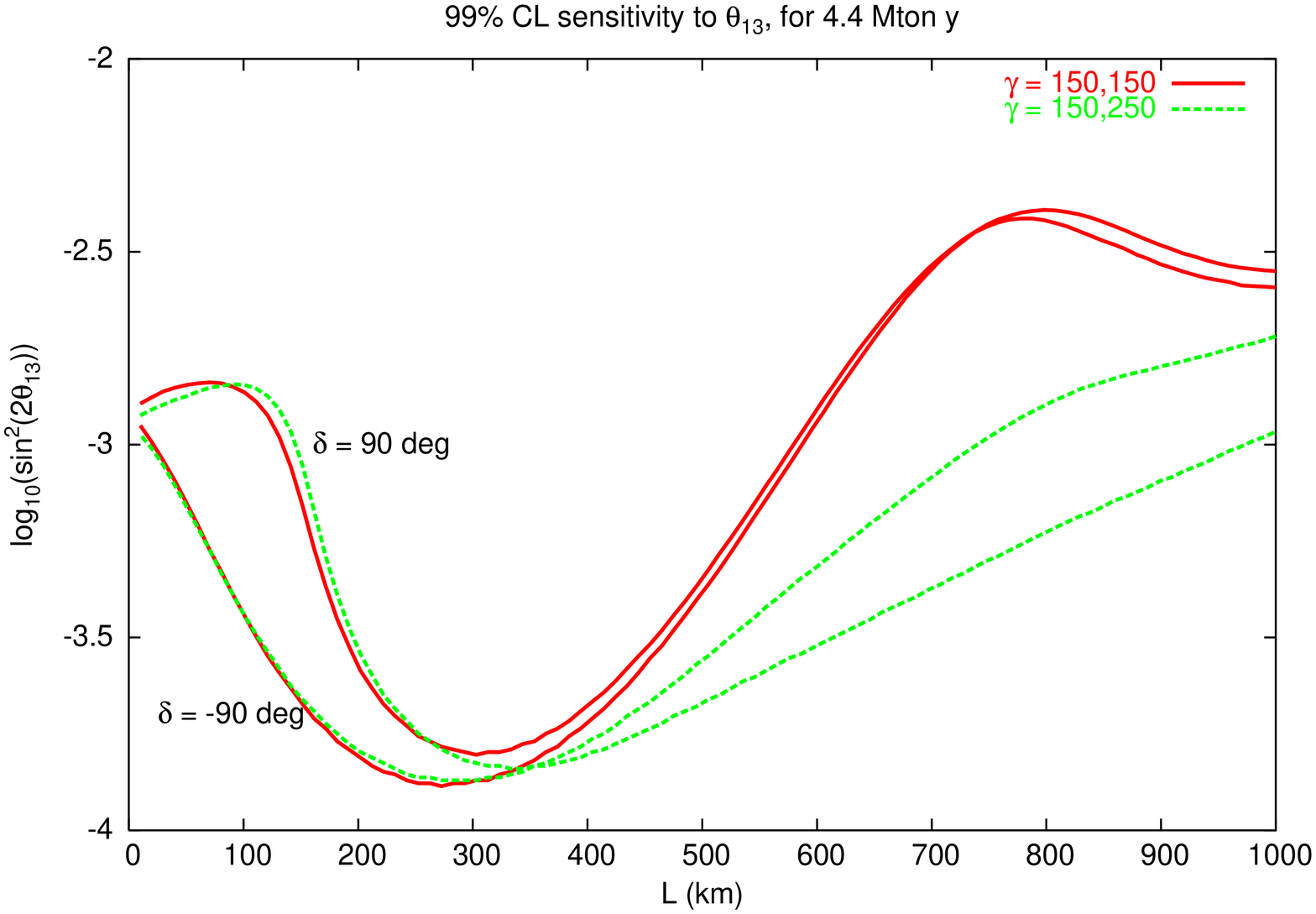,width=7cm,height=7cm,angle=-90} 

\caption{\it Left: minimum value of $|\delta|$ distinguishable from $0$ and $180^\circ$ at 99\%~CL (for $\theta_{13}=8^\circ$) vs. baseline for $\gamma=150/150$ (red) and $\gamma=150/250$ (green). Right: minimum value of $\theta_{13}$ distinguishable from $0$ at 99\%~CL (for $\delta=90^\circ$ and $-90^\circ$ as shown).}  
\la{fig:versusL}

\end{center}
\end{figure}

Setup II will hence be defined as $\gamma=150/150$ for $L=300$~km. 
Similar results are expected for the asymmetric option $\gamma=150/250$ with slightly longer baseline. 

\section{Comparison of the three setups}

From the results of the previous section, the default setups be compared are:
\begin{itemize}
\item Setup I: $\gamma_{\helio}=\gamma_{\neon}=120$ at $L=130$~km
\item Setup II: $\gamma_{\helio}=\gamma_{\neon}=150$ at $L=300$~km
\item Setup III: $\gamma_{\helio}=\gamma_{\neon}=350$ at $L=730$~km
\end{itemize}
For the highest $\gamma$ option, we have also checked that the symmetric and asymmetric options give comparable results. 

%
\subsection{Intrinsic sensitivity to $\theta_{13}$ and $\delta$}

Figure~\ref{fig:exc_int} compares the CP-violation 
and $\theta_{13}$ exclusion plots for the three setups assuming, for the moment, that the discrete ambiguities in $\sigdma$ and $\sigta$ can be ignored because correct assignments have been made.  Also included for reference is
the previously considered setup from~\cite{blm}. Although the 
highest $\gamma$ option of~\cite{bbeam} remains best, the 
performance of Setup II is comparable. Even the sensitivity of the 
much-improved
CERN--Frejus scenario in Setup I is considerable. Although only the range $(-90^\circ,90^\circ)$ is shown, to make it easier to read the y-scale, the region around $180^\circ$ has a similar pattern.

As explained in~\cite{bbeam}, differences between the setups arise due to 
sample size (which increases at least linearly with $\gamma$) and more robust energy reconstruction at higher energies (as Fermi motion becomes less important).




\begin{figure}[t]
\begin{center}

\epsfig{file=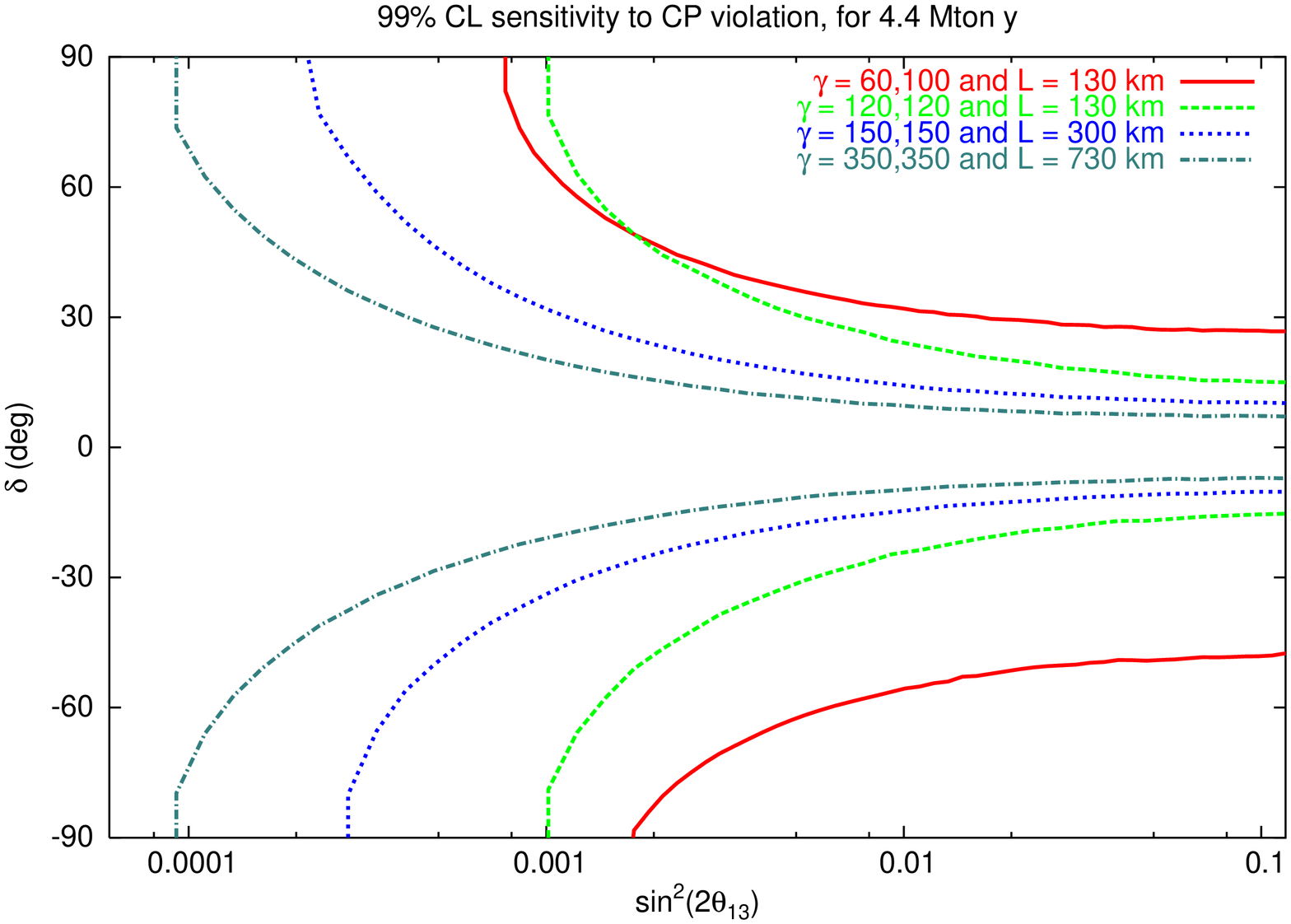,width=7cm,height=7cm,angle=-90} 
\epsfig{file=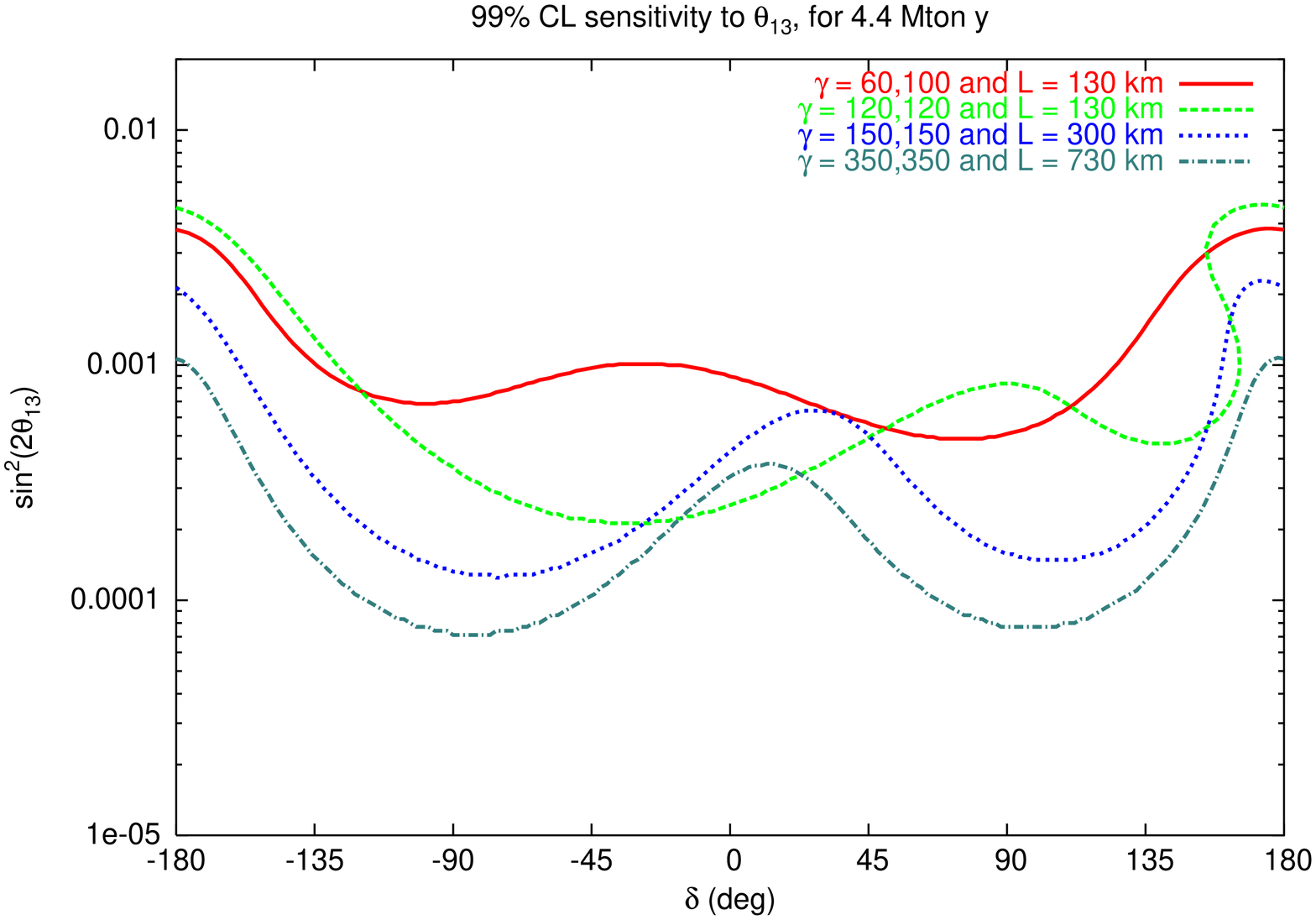,width=7cm,height=7cm,angle=-90} 

\caption{\it Left: CP-violation exclusion plot at 99\%~CL for the three
reference Setups I (dashed), II (dotted) and III (dashed-dotted) compared with the {\it standard} (solid) one of~\cite{mauro,blm}. Right: exclusion plot for $\theta_{13}$ at 99\%~CL with the same setups. 
The solar and atmospheric parameters are fixed to their present best fit values and 
the discrete ambiguities are assumed to be resolved.}
\la{fig:exc_int}
\end{center}
\end{figure}

Figure~\ref{fig:fit_int} shows typical fits for the three setups at 
several {\it true} values of $\theta_{13}$ and $\delta$. While both Setups II
and III manage to resolve the intrinsic degeneracy essentially everywhere in the sensitivity range,
this is not the case for Setup I; there (when the fake solution gets closer to and merges with the true one) the errors in $\theta_{13}$ and $\delta$ are sometimes strongly enhanced by the intrinsic degeneracy. This effect is not 
necessarily noticeable in the exclusion plot for CP violation.

\begin{figure}[t]
\begin{center}

\epsfig{file=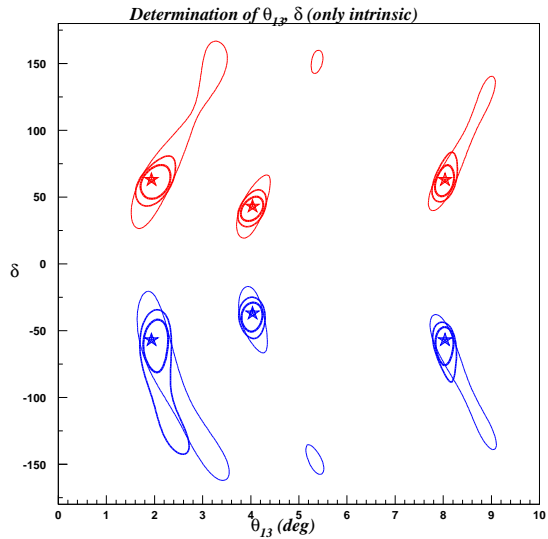,width=8cm,height=8cm} 

\caption{\it Determination of $(\theta_{13},\delta)$ at 99\%~CL for Setups
III (thicker line), II (intermediate) and I (thinner line) and six different true 
values of the parameters indicated by the stars, assuming the 
correct $\sigdma$ and $\sigta$.}
\la{fig:fit_int}
\end{center}
\end{figure}

\subsection{Effect of the eight-fold degeneracies}

By the time any $\beta$-beam begins, it is probable that 
a number of uncertainties in the oscillation parameters besides $\theta_{13}$
and $\delta$ will remain, in particular the discrete ambiguity in $\sigdma$ or the octant of $\theta_{23}$. Both questions
are theoretically important and the possibility of answering them with a
$\beta$-beam is attractive. 
These ambiguities are problematic, if they can't be resolved, because they 
can bias the determination of the parameters $(\theta_{13}, \delta)$, that is,
the solutions surviving with the wrong assignment of the sign and/or the
octant lie at different values of $\theta_{13}$ and $\delta$ than the true ones. 

Generically, an eight-fold degeneracy of solutions
appears when only the golden channel is measured and no energy dependence is available. There are two solutions
in the absence of the discrete ambiguities, the true and the intrinsic one~\cite{burguet1}.
Each gets an false image for the wrong assignment of the sign~\cite{mn}, for the octant~\cite{fl,bargerdeg}
and for both. 

As explained in~\cite{burguet2}, the intrinsic solution and its three
images are strongly dependent on the neutrino energy and therefore can 
be excluded, in principle, when the energy dependence of the oscillation signal
is significant. On the other hand images of the true solution are 
energy independent and impossible to resolve unless there are
additional measurements (e.g. disappearance measurements or the silver channel),
or when there are significant matter effects.

\begin{figure}[t]
\begin{center}

\epsfig{file=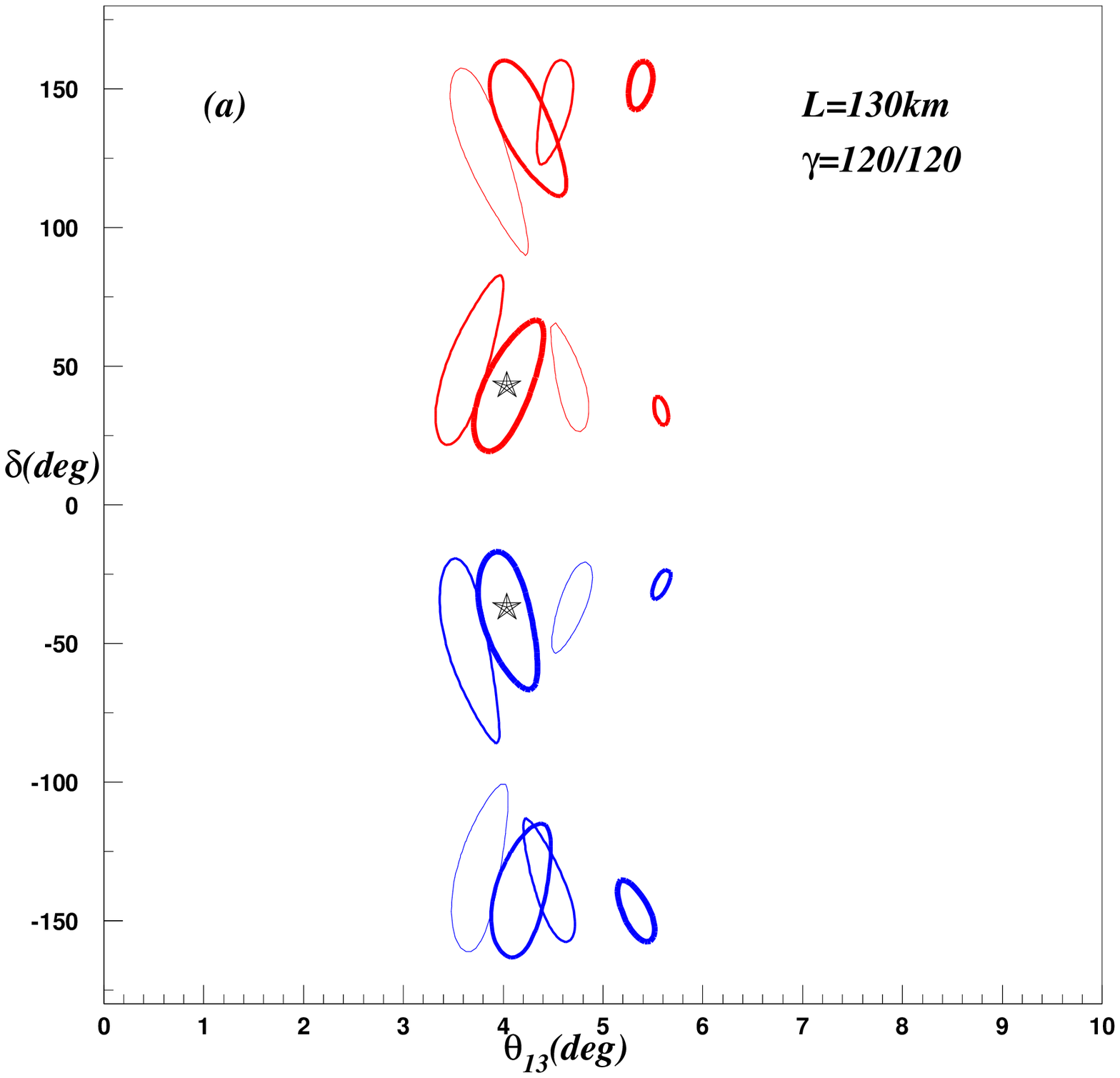,width=9cm,height=7cm}

 \epsfig{file=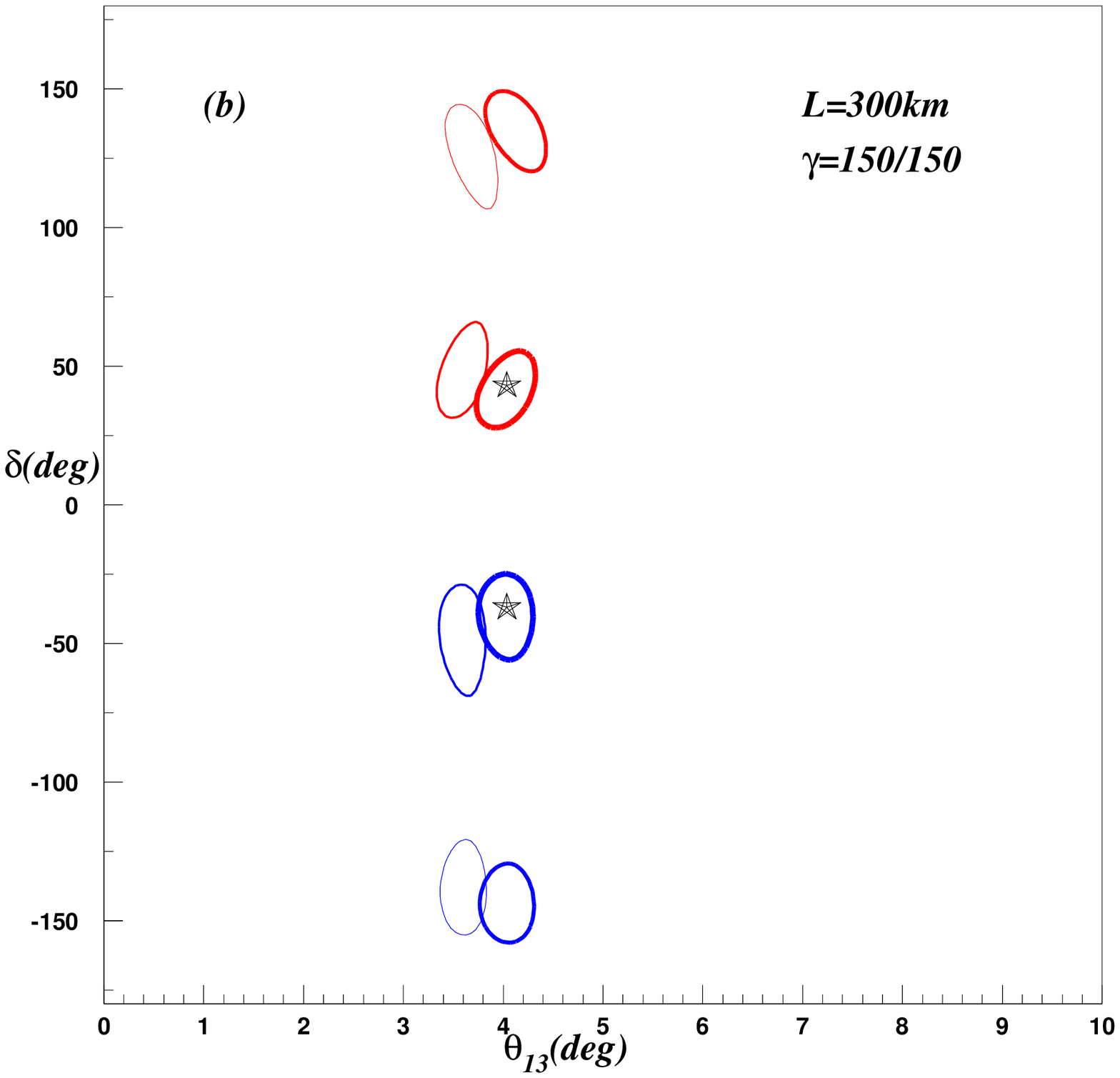,width=9cm,height=7cm}

 \epsfig{file=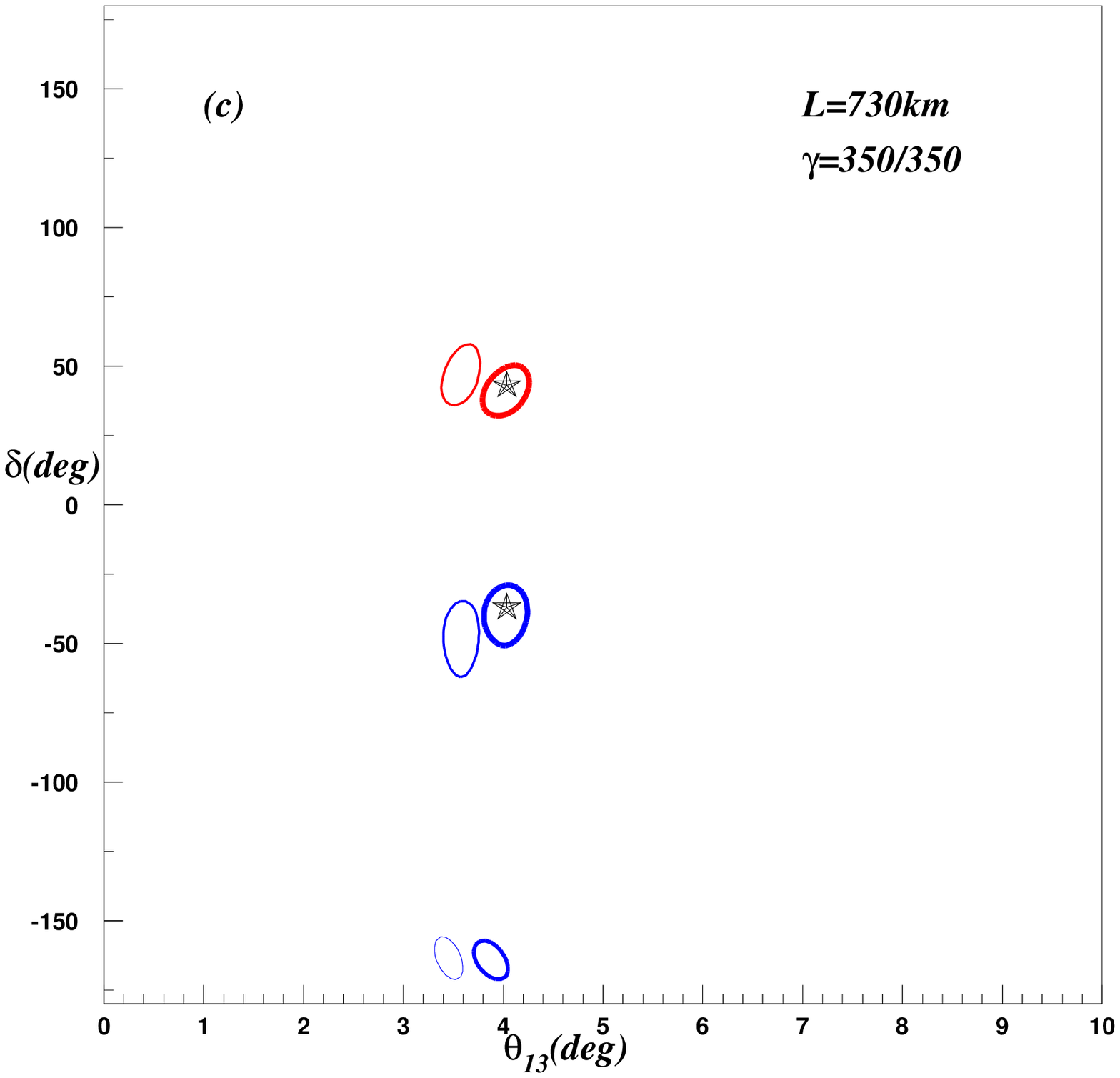,width=9cm,height=7cm} 

\caption{\it Solutions for $(\theta_{13},\delta)$ for the true values: $\delta=\pm 40^\circ$ and $\theta_{13}=4^\circ$ in (a) Setup I, (b) Setup II and (c) Setup III without discrete ambiguities, with the sign, octant and mixed ambiguities ordered from thicker to thinner-line contours.} 
\la{fig:fit_deg}
\end{center}
\end{figure}

Figure~\ref{fig:fit_deg} shows fits including the discrete ambiguities on the plane $(\theta_{13}, \delta)$ for the three
setups and different choices of the true $\theta_{13}$ and $\delta$.
In Setup I we generically find the full eight-fold degeneracy, while 
in Setups II and III the intrinsic solution and its images are typically 
excluded, thanks to the stronger energy dependence.

Some general observations concerning these results include:
\begin{itemize}
\item Presence of the intrinsic degenerate solution or its images as in Setup I is 
problematic, because it implies a significant increase in the 
measurement errors of $\theta_{13}$ and $\delta$ (as shown in Figure~\ref{fig:fit_int}) for some values of $\delta$). 
\item When only the images of the true solution survive, as in Setups II and III, they 
interfere with the measurement of $\theta_{13}$ and $\delta$ by 
mapping the true solution to another region of parameter space. In vacuum~\cite{mn,burguet2}:

Wrong-sign: $\theta_{13} \rightarrow \theta_{13}, \delta \rightarrow \pi -\delta$

Wrong-octant: $\theta_{13} \rightarrow \tan \theta_{23} \theta_{13} + {\cal O}(\Delta m^2_{12}),\;\;\; \sin \delta \rightarrow \cot \theta_{23} \sin\delta$

Since these different regions occur for different choices of the discrete
ambiguities they cannot overlap and one ends with a set of distinct
measurements of $\theta_{13},\delta$ with different
central values but similar errors (see the middle and right plots of Figure~\ref{fig:fit_deg}). 

\item In vacuum, CP violating solutions are mapped into CP violating solutions, therefore the effects of degeneracies on the exclusion plot for CP violation are often small, even when degeneracies are a problem. 
In matter, on the other hand, $\delta$ shifts
in the fake solutions are enhanced by matter effects and 
for some central values of $(\theta_{13},\delta)$ the fake solutions  
may move closer to the CP 
conserving lines than the true solution, resulting in an apparent loss of 
sensitivity to CP violation. This effect is visible in Figure~\ref{fig:fit_deg} where
the fake-sign solution, which 
in vacuum should be located at $\sim -140^\circ$ for $\delta=-40^\circ$, gets
shifted towards the CP conserving line$-180^\circ$ for longer baselines where 
matter effects are larger. 

\end{itemize}

Figures~\ref{fig:exc_sign} and~\ref{fig:exc_oct} show the range of $(\theta_{13},\delta)$, where the $\sigdma$ and 
$\sigta$ can be measured respectively. Asymmetric $\gamma$ options are also included, since there are some differences.
As expected, sensitivity to the discrete ambiguities is better for large $\theta_{13}$ and larger $\gamma$.
In Setup I there is essentially no sensitivity anywhere on the plane.

\begin{figure}[t]
\begin{center}

\epsfig{file=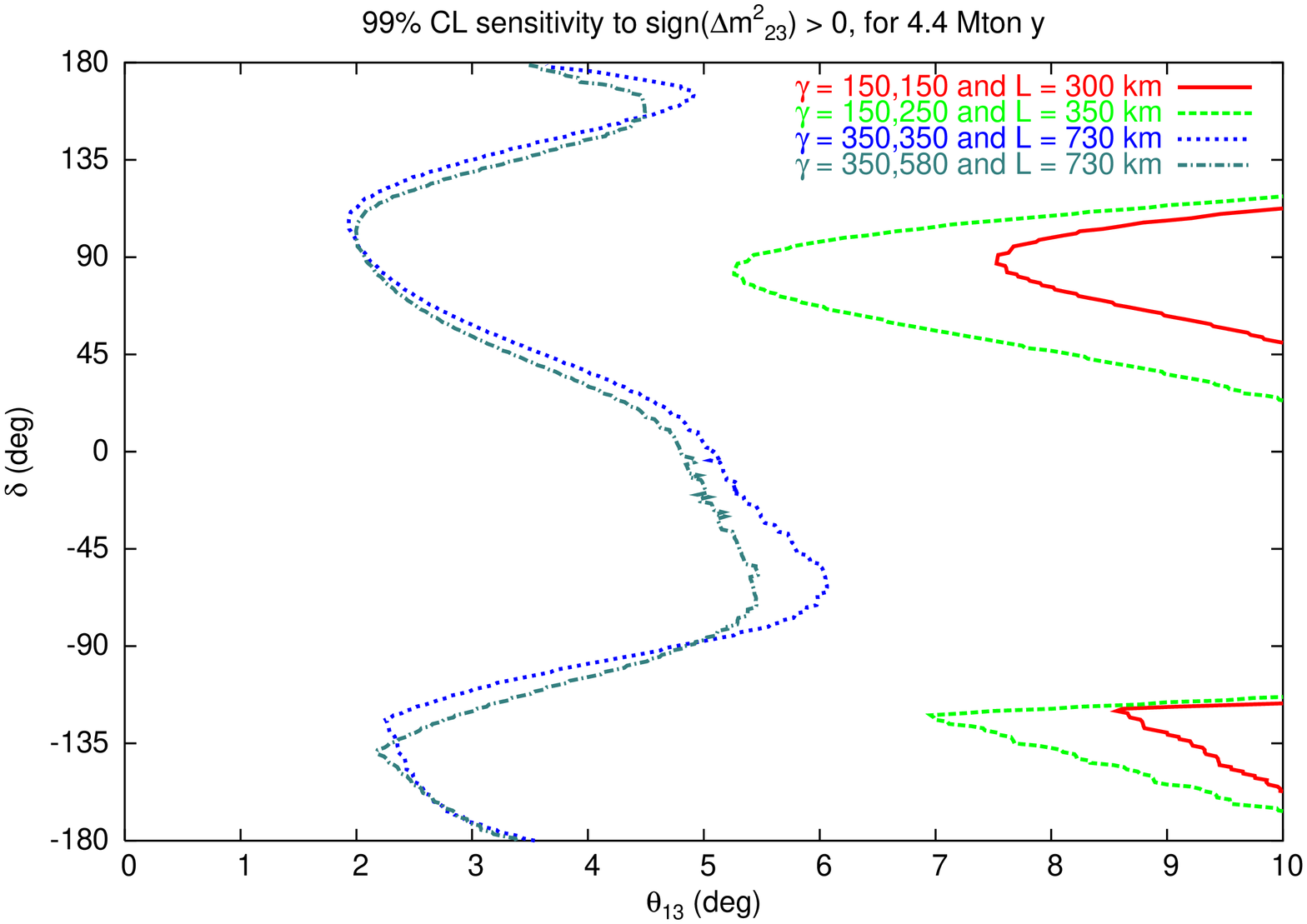,width=7cm,height=7cm,angle=-90} 
\epsfig{file=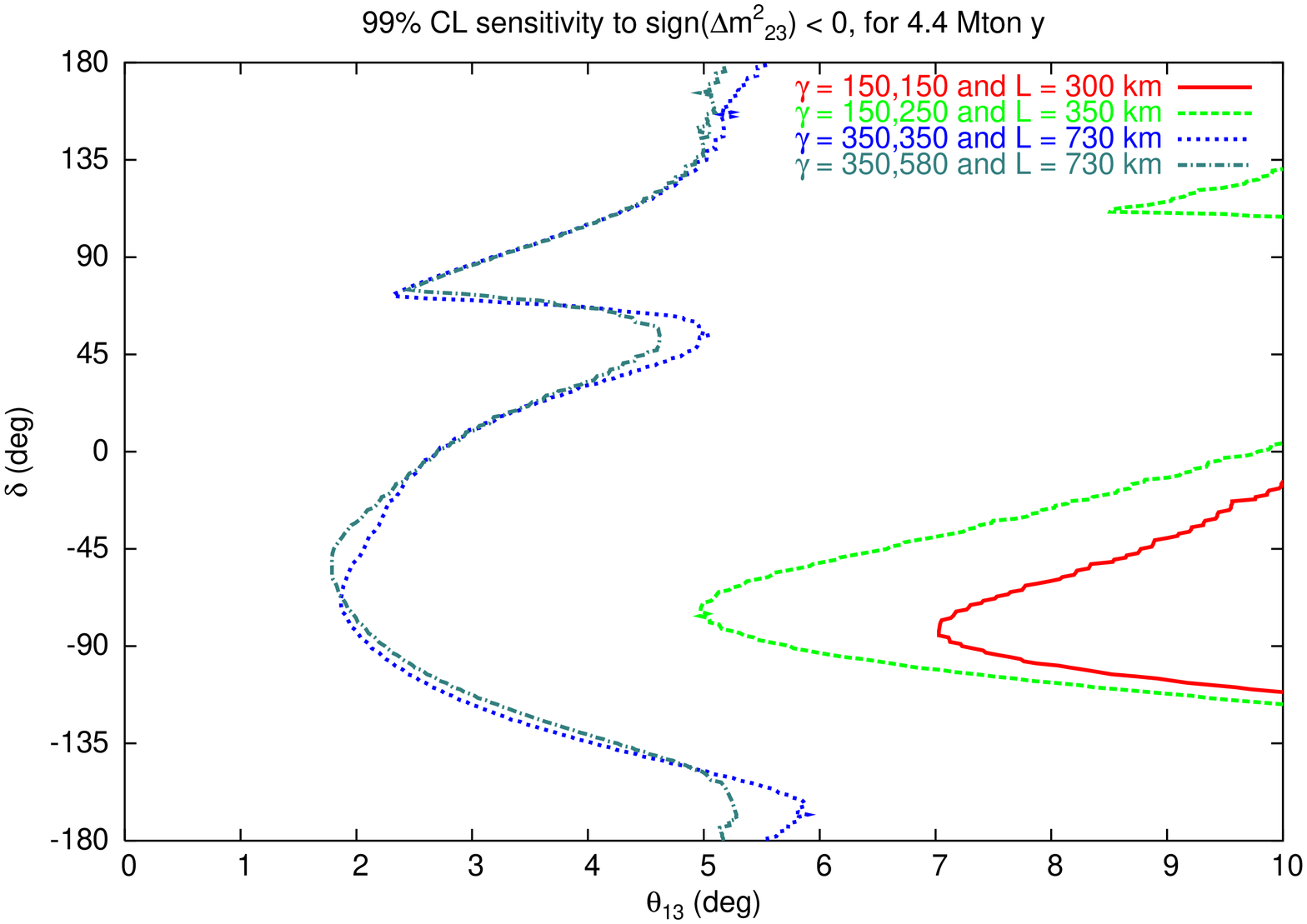,width=7cm,height=7cm,angle=-90} 

\caption{\it Region on the plane $(\theta_{13},\delta)$ in which $\sigdma$ can be measured at 99\%~CL for $\theta_{23}=40.7^\circ$ and positive (left) and negative (right) $\delta m^2_{23}$.  Symmetric and asymmetric beam options are shown for Setup II (300~km, solid and dashed, respectively) and Setup III (730~km, dotted and dash-dot).  There is no sensitivity for Setup I.}
\la{fig:exc_sign}
\end{center}
\end{figure}

\begin{figure}[t]
\begin{center}

\epsfig{file=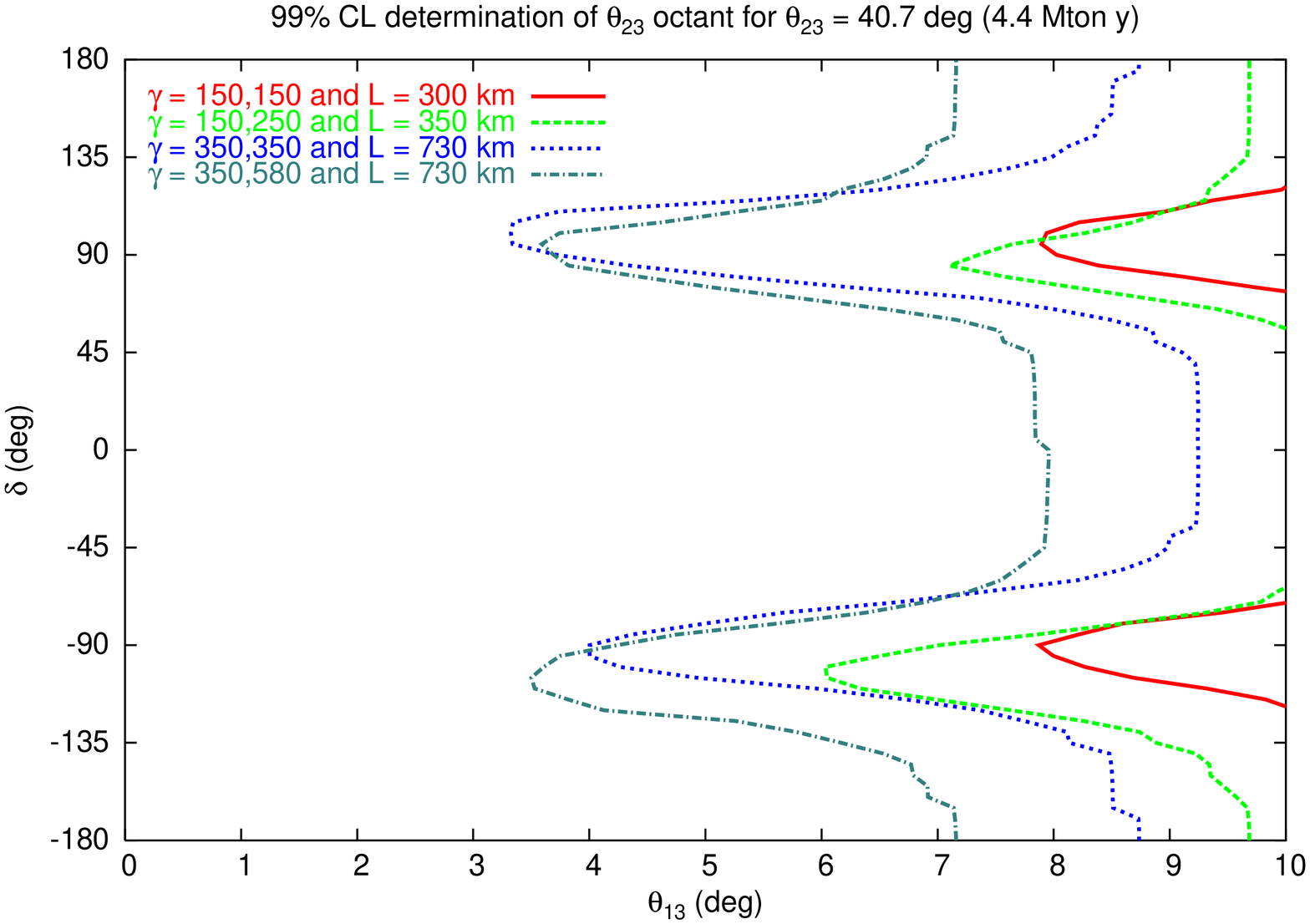,width=7cm,height=7cm,angle=-90} 
\epsfig{file=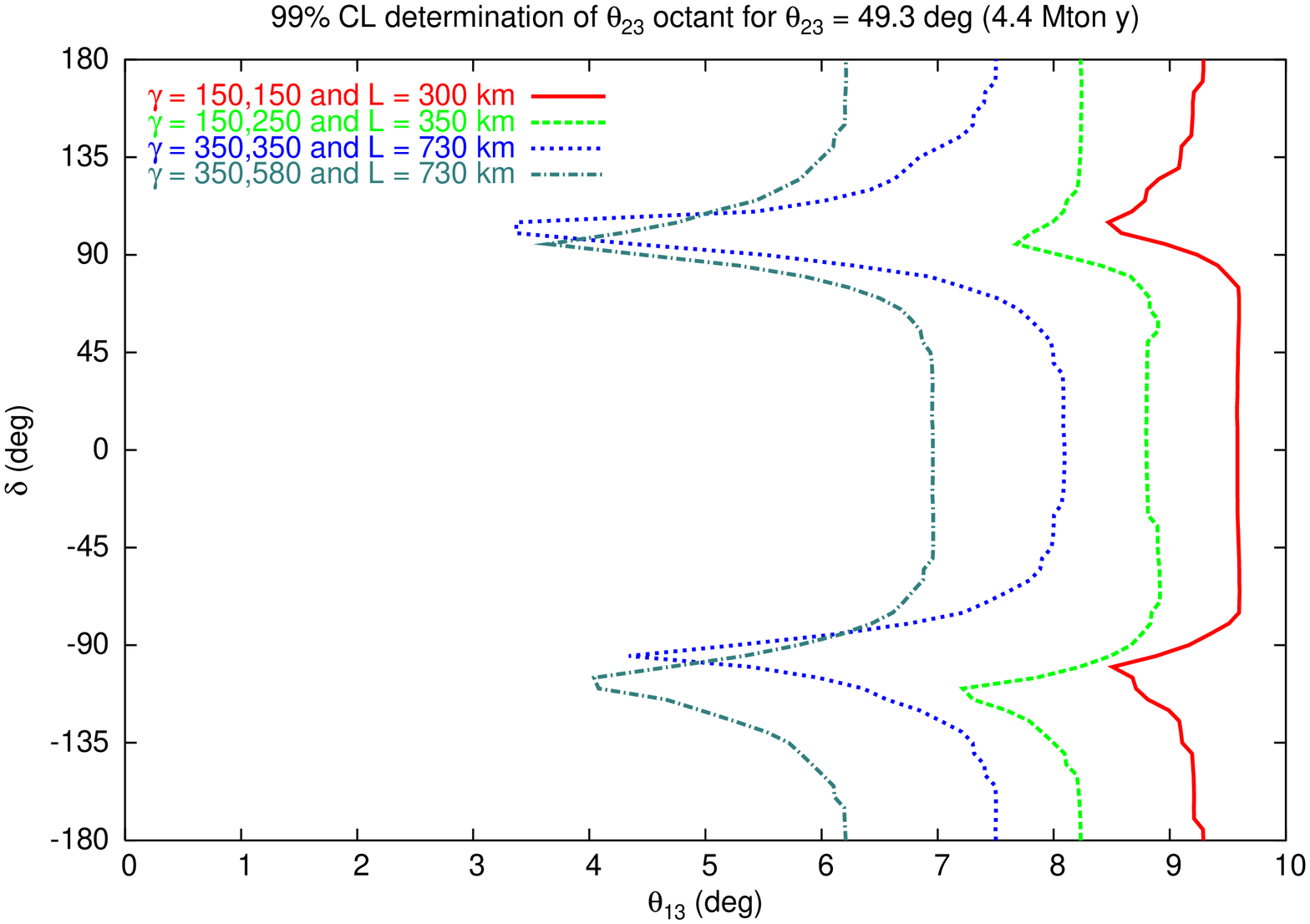,width=7cm,height=7cm,angle=-90} 

\caption{\it Region on the plane $(\theta_{13},\delta)$ in which $\sigta$ can be measured at 99\%~CL for 
$\theta_{23}=40.7^\circ$ (left) and $\theta_{23}=49.3^\circ$ (right). Setups II (300~km, solid: symmetric, dashed: asymmetric) and  III (730~km, dotted: symmetric, dash-dot: asymmetric) are shown.  There is no sensitivity for Setup I.}
\la{fig:exc_oct}
\end{center}
\end{figure}

Sensitivity to the discrete
ambiguities and their bias in the determination of the parameters
$\theta_{13}$ and $\delta$ could be significantly improved if  
data for any of the setups is combined with 
$\nu_\mu\rightarrow\nu_\mu$ disappearance measurements, for instance in a superbeam experiment.  This combination was recently studied in~\cite{doninidis} for the {\it standard} $\beta$-beam with significant improvement in sensitivity to the mass hierarchy, even without matter effects.
 A similar study for the setups considered here will be very interesting. One of the most important limitations of the $\beta$-beam, 
compared to the superbeam or the Neutrino Factory, is its inability to
measure the atmospheric parameters $(\theta_{23}, \Delta m^2_{23})$ with precision.
At the very least, information from T2K phase-I should be included, since otherwise the uncertainty on these
parameters will seriously compromise sensitivity to $\theta_{13}$ and $\delta$.  Synergies in resolving degeneracies, between the $\beta$-beam and T2K, should also be exploited. 

Another interesting observation is that atmospheric neutrinos can be measured in the same megaton detector considered here. A recent study~\cite{lblatm} combining atmospheric data with T2K phase-II has found a large improvement in sensitivity of the latter to both discrete ambiguities when $\theta_{13}$ is not too small ($> 4^\circ$). This is surely an analysis that should be done and will be reported elsewhere.

\section{Conclusions}

This paper has explored the physics potential of a CERN-SPS $\beta$-beam, where ions can be accelerated to $\gamma_{\helio}\leq 150$ and  $\gamma_{\neon}\leq 250$.  The design of a $\beta$-beam reaching this maximum $\gamma$ is technically equivalent 
to the lower-$\gamma$ option previously considered, for which a feasibility study already exists~\cite{bbcern}.  A major improvement in sensitivity to $\theta_{13}$ and $\delta$ is achieved by increasing $\gamma$.  Even 
when the baseline is fixed to that of CERN--Frejus, sensitivity improves considerably if $\gamma > 100$ and changes slowly as the $\gamma$ increases further to the limit of the SPS.  

An even more dramatic improvement is possible if the baseline is increased proportionally, so the first atmospheric oscillation
maximum corresponds to the average neutrino energy, which occurs at $L\sim 300$~km. For large values of $\theta_{13}$ 
this option is comparable in CP violation sensitivity to the optimal one in~\cite{bbeam} at even higher $\gamma \sim {\cal O}(400)$, which would require a more powerful accelerator, such as the Tevatron or a refurbished SPS. In contrast, for small values of $\theta_{13}$ the latter option is still significantly better.

The main differences can be traced to increased
event rate and the more significant energy dependence, which allows higher-$\gamma$ options to resolve the intrinsic degeneracy. 

For discrete ambiguities, higher-$\gamma$ 
also provides a window on the neutrino mass hierarchy and the octant
of $\theta_{23}$, if non-maximal, relying on 
significant matter effects; the highest-$\gamma$ setup with 
$730$~km baseline is therefore the only one with a significant sensitivity. 

In summary, if the existing CERN-SPS is the ion accelerator and the CERN--Frejus baseline is fixed, 
$\gamma$ should still be increased to a value greater than 100, higher than considered in~\cite{mauro,blm}. 
If an alternative site hosts a large underground laboratory near CERN, it will be profitable to exploit longer baselines
$L=300$~km. In any case, R\&D effort to design $\beta$-beams beyond the limit of the CERN-SPS appears justified, given the 
significant improvements in physics sensitivity they would allow.

\section*{Acknowledgement}

 We wish to thank A.~Blondel, A.~Donini, E.~Fern\'andez-Mart\'{\i}nez, M.B.~Gavela, M.~Mezzetto and S.~Rigolin for useful discussions.    
This work has been partially supported 
by  CICYT (grants FPA2002-00612, FPA-2003-06921, FPA2004-00996), Generalitat Valenciana (GV00-054-1, GV2004-B-159), CARE-BENE (European Integrated Activity) and by the
U.S. Department of Energy grant DE-FG02-91ER40679.


\appendix
\renewcommand{\thesection}{Appendix~\Alph{section}}
\renewcommand{\thesubsection}{\Alph{section}.\arabic{subsection}}
\renewcommand{\theequation}{\Alph{section}.\arabic{equation}}


\end{document}